\documentclass{aa}
\usepackage{color}
\usepackage{graphicx}
\usepackage{lscape}
\usepackage{natbib}
\usepackage[scaled]{helvet}
\usepackage[varg]{txfonts}
\usepackage{url}
\usepackage{xspace}
\bibpunct{(}{)}{;}{a}{}{,}
\newcounter{Rco}
\newcommand{\ionw}[3]{\mbox{\ion{#1}{#2}~$\lambda\,#3\,\mathrm{\AA}$}\xspace}
\newcommand{\ionww}[3]{\mbox{\ion{#1}{#2}~$\lambda\lambda\,#3\,\mathrm{\AA}$}\xspace}

\newcommand{\logg}{\mbox{$\log g$}\xspace}
\newcommand{\loggw}[1]{\mbox{$\log g\hspace{-0.5mm} =\hspace{-0.5mm}  #1$}}

\newcommand{\sla}{\raisebox{-0.10em}{$\stackrel{<}{{\mbox{\tiny $\sim$}}}$}}

\newcommand{\Teff}{\mbox{$T_\mathrm{eff}$}\xspace}
\newcommand{\Teffw}[1]{\mbox{$\Teff\hspace{-0.5mm} =\hspace{-0.5mm} #1 \,\mathrm{K}$}}

\newcommand{\mmspr}{\hbox{}\hspace{+0.7cm}}

\newcommand{\gb}{\object{G191$-$B2B}\xspace}
\newcommand{\re}{\object{RE\,0503$-$289}\xspace}
\begin{document}

\title{Stellar laboratories}
\subtitle{VI. New \ion{Mo}{iv -- vii} oscillator strengths and the molybdenum abundance \\
              in the hot white dwarfs \gb and \re
           \thanks
           {Based on observations with the NASA/ESA Hubble Space Telescope, obtained at the Space Telescope Science 
            Institute, which is operated by the Association of Universities for Research in Astronomy, Inc., under 
            NASA contract NAS5-26666.
           }$^,$
           \thanks
           {Based on observations made with the NASA-CNES-CSA Far Ultraviolet Spectroscopic Explorer.
           }$^,$
           \thanks
           {Tables 11 to 14 are only available via the
            German Astrophysical Virtual Observatory (GAVO) service TOSS (http://dc.g-vo.org/TOSS).
           }
         }
\titlerunning{Stellar laboratories: new \ion{Mo}{iv -- vii} oscillator strengths}

\author{T\@. Rauch\inst{1}
        \and
        P\@. Quinet\inst{2,3}
        \and
        D\@. Hoyer\inst{1}
        \and
        K\@. Werner\inst{1}
        \and
        M\@. Demleitner\inst{4}
        \and
        J\@. W\@. Kruk\inst{5}
        }

\institute{Institute for Astronomy and Astrophysics,
           Kepler Center for Astro and Particle Physics,
           Eberhard Karls University,
           Sand 1,
           72076 T\"ubingen,
           Germany \\
           \email{rauch@astro.uni-tuebingen.de}
           \and
           Physique Atomique et Astrophysique, Universit\'e de Mons -- UMONS, 7000 Mons, Belgium
           \and
           IPNAS, Universit\'e de Li\`ege, Sart Tilman, 4000 Li\`ege, Belgium
           \and
           Astronomisches Rechen-Institut, Zentrum f\"ur Astronomie, Ruprecht Karls University, M\"onchhofstra\ss e 12-14, 69120 Heidelberg, Germany
           \and
           NASA Goddard Space Flight Center, Greenbelt, MD\,20771, USA}

\date{Received 8 September 2015; accepted 15 December 2015}

\abstract {For the spectral analysis of high-resolution and high signal-to-noise (S/N) spectra of hot stars,
           state-of-the-art non-local thermodynamic equilibrium (NLTE) 
           model atmospheres are mandatory. These are strongly
           dependent on the reliability of the atomic data that is used for their calculation.
          }
          {To identify molybdenum lines in the ultraviolet (UV) spectra of
           the DA-type white dwarf \gb and
           the DO-type white dwarf \re
           and, to determine their photospheric Mo abundances, reliable  
           \ion{Mo}{iv-vii} oscillator strengths are used.
          }
          {We newly calculated \ion{Mo}{iv-vii} oscillator strengths
           to consider their radiative and collisional bound-bound transitions
           in detail in our NLTE stellar-atmosphere models
           for the analysis of Mo lines exhibited in
           high-resolution and high S/N UV observations of \re.
          }
          {We identified 12 \ion{Mo}{v} and nine \ion{Mo}{vi} lines in the UV spectrum of \re and
           measured a photospheric Mo abundance of $1.2 - 3.0\,\times\,10^{-4}$ (mass fraction, 22\,500\,$-$\,56\,400 times the solar abundance).
           In addition, from the \ion{As}{v} and \ion{Sn}{iv} resonance lines, we measured mass fractions 
           of arsenic ($0.5 - 1.3\,\times\,10^{-5}$, about 300\,$-$\,1200 times solar) and
           tin  ($1.3 - 3.2\,\times\,10^{-4}$, about 14\,300\,$-$\,35\,200 times solar).
           For \gb, upper limits were determined for the abundances of Mo ($5.3\times 10^{-7}$, 100 times solar) and, in addition, for
           Kr ($1.1\times 10^{-6}$, 10 times solar) and
           Xe ($1.7\times 10^{-7}$, 10 times solar).
           The arsenic abundance was determined ($2.3 - 5.9\,\times\,10^{-7}$, about 21\,$-$\, 53 times solar).
           A new, registered German Astrophysical Virtual Observatory (GAVO) service, TOSS, has been constructed to provide
           weighted oscillator strengths and transition probabilities.
           }
          {Reliable measurements and calculations of atomic data are a prerequisite for
           stellar-atmosphere modeling. 
           Observed \ion{Mo}{v-vi} line profiles in the UV spectrum of the white dwarf \re 
           were well reproduced with our newly calculated oscillator strengths. 
           For the first time, this allowed the photospheric Mo abundance in a white dwarf to be determined.
          }


\maketitle

\section{Introduction}
\label{sect:intro}

\re \citep[\object{WD\,0501$-$289},][]{mccooksion1999,mccooksion1999cat} is a hot, helium-rich, DO-type white dwarf
\citep[WD, effective temperature \Teffw{70\,000}, surface gravity $\log\,(g\,/\,\mathrm{cm/s^2}) = 7.5$,][]{dreizlerwerner1996},
that exhibits lines of at least ten trans-iron elements in its far-ultraviolet (FUV) spectrum \citep{werneretal2012}.
The abundance analysis of these species is hampered by the lack of atomic data for their
higher ionization stages, i.e., \ion{}{iv-vii}. While \citet{werneretal2012} could measure only the Kr and Xe abundances,
further abundance determinations
\citep[Zn, Ge, Ga, Xe, and Ba by][respectively]{rauchetal2014zn,rauchetal2012ge,rauchetal2014ba,rauchetal2015ga,rauchetal2015xe}
were always initiated by new calculations of reliable transition probabilities.

\gb \citep[\object{WD\,0501+527},][]{mccooksion1999,mccooksion1999cat} is a hot, hydrogen-rich, DA-type white dwarf that was
recently analyzed by 
\citet[][\Teffw{60\,000}, \loggw{7.6}]{rauchetal2013}. 
Based on this model,
\citet{rauchetal2014zn,rauchetal2014ba,rauchetal2015ga} measured the abundances of Zn, Ba, and Ga.

Molybdenum is another trans-iron element (atomic number $Z = 42$). It was discovered for the first time in a WD (in the spectrum of \re) by 
\citet[][four \ion{Mo}{vi} lines]{werneretal2012}.
To identify more lines of Mo and to determine its abundance, we calculated new transition probabilities for \ion{Mo}{iv-vii}.

In this paper, we first describe the available observations (Sect.\,\ref{sect:observation}),
our stellar-atmosphere models (Sect.\,\ref{sect:models}), and
the computation of the new transition probabilities (Sect.\,\ref{sect:motrans}).
A new Virtual Observatory (VO) service that provides access to transition probabilities is
presented in Sect.\,\ref{sect:toss}.

To use the most elaborated models of 
\gb and \re
for our Mo abundance analysis, we start with an incorporation and an abundance determination of 
arsenic (Sect.\,\ref{sect:as}, both stars) and 
tin  (Sect.\,\ref{sect:snre}, \re). 
Then, we assess the Mo photospheric abundances in 
\re and \gb (Sect.\,\ref{sect:more}). 
In Sect.\,\ref{sect:krxegb}, we determine upper abundance limits of krypton and xenon in \gb.
To understand the abundance patterns of trans-iron elements,
we investigate on the efficiency of radiative levitation acting on the elements 
Zn, Ga, Ge, As, Kr, Mo, Sn, Xe, and Ba in both stars' atmospheres (Sect.\,\ref{sect:diffusion}).
We summarize our results and conclude in Sect.\,\ref{sect:results}.

\section{Observations}
\label{sect:observation}

\paragraph{\gb}
We used the spectra obtained with the Far Ultraviolet Spectroscopic Explorer 
\citep[FUSE,
       $910\,\mathrm{\AA} < \lambda <  1190\,\mathrm{\AA}$, 
       resolving power $R = \lambda/\Delta\lambda \approx 20\,000$, for details see]
      []
       {rauchetal2013}
       and the Hubble Space Telescope / Space Telescope Imaging Spectrograph 
\citep[HST/STIS,
       $1145\,\mathrm{\AA} < \lambda <  3145,\mathrm{\AA}$, resolution of $\approx\,3$\,km/s, see]
      [available at \url{http://www.stsci.edu/hst/observatory/cdbs/calspec.html}
      ]{rauchetal2013}.

\paragraph{\re}
We analyzed its FUSE 
\citep[described in detail by][]{werneretal2012}
and HST/STIS observations ($1144\,\mathrm{\AA} < \lambda <  3073\,\mathrm{\AA}$). The latter was co-added from 
two observations with grating E140M (exposure times 2493\,s and 3001\,s, $1144\,\mathrm{\AA} - 1709\,\mathrm{\AA}$, $R \approx 45\,800$), and
two observations with grating E230M (1338\,s, $1690\,\mathrm{\AA} - 2366\,\mathrm{\AA}$ and 
                                     1338\,s, $2277\,\mathrm{\AA} - 3073\,\mathrm{\AA}$, $R \approx 30\,000$).
These STIS observations are retrievable from the Barbara A\@. Mikulski Archive for Space Telescopes (MAST).

\section{Model atmospheres and atomic data}
\label{sect:models}

We employed the  T\"ubingen NLTE\footnote{non-local thermodynamic equilibrium} Model Atmosphere Package
\citep[TMAP\footnote{\url{http://astro.uni-tuebingen.de/~TMAP}},][]{werneretal2003,tmap2012}
to calculate plane-parallel, chemically homogeneous model-atmospheres in hydrostatic and radiative 
equilibrium. 
Model atoms were taken from the T\"ubingen Model Atom Database
\citep[TMAD\footnote{\url{http://astro.uni-tuebingen.de/~TMAD}},][]{rauchdeetjen2003} that
has been constructed as part of the T\"ubingen contribution to the German Astrophysical Virtual Observatory 
(GAVO\footnote{\url{http://www.g-vo.org}}).
For our Mo model atoms, we follow \citet{rauchetal2015ga} and used a statistical approach to calculate so-called
super levels and super lines with our Iron Opacity and Interface
\citep[IrOnIc\footnote{\url{http://astro.uni-tuebingen.de/~TIRO}},][]{rauchdeetjen2003}. 
We transferred our new Mo data into Kurucz-formatted files\footnote{GFxxyy.GAM, GFxxyy.LIN, and GFxxyy.POS 
files with xx = element number, yy = element charge, \url{http://kurucz.harvard.edu/atoms.html}}
that were then ingested and processed by IrOnIc. 
The statistics of our Mo model atom is summarized in Table\,\ref{tab:ironic}.

\begin{table}\centering
\caption{Statistics of \ion{Mo}{iv - vii} atomic levels and line transitions from
         Tables\,11 - 14, respectively.}         
\label{tab:ironic}
\begin{tabular}{ccccc}
\hline
\hline
ion       & Atomic levels & Lines & Super levels & Super lines \\
\hline
\sc{iv}   &           162 &  2803 &            7 &          15 \\
\sc{v}    &           257 &  5882 &            7 &          22 \\
\sc{vi}   &           112 &   988 &            7 &          23 \\
\sc{vii}  &            95 &  1181 &            7 &          16 \\
\hline
          &           626 & 10824 &           28 &          76 \\
\hline
\end{tabular}
\end{table}  

For Mo and all other species, level dissolution (pressure ionization) following
\citet{hummermihalas1988} and \citet{hubenyetal1994} is accounted for. 
Broadening for all Mo lines that are due to the quadratic Stark effect is calculated 
using approximate formulae by \citet{cowley1970,cowley1971}.

\section{Atomic structure and radiative data calculation}
\label{sect:motrans}

New calculations of oscillator strengths for a large number of transitions of molybdenum ions that are considered 
in the present work were carried out using the pseudo-relativistic Hartree-Fock (HFR) method 
\citep{cowan1981}, including core-polarization corrections \citep[see, e.g.,][]{quinetetal1999,quinetetal2002}.

For \ion{Mo}{iv}, the configuration interaction was considered among the configurations 
4d$^3$,  
4d$^2$5s, 
4d$^2$6s, 
4d$^2$5d, 
4d$^2$6d, 
4d4f$^2$, 
4d5s$^2$, 
4d5p$^2$, 
4d5d$^2$, 
4d5s5d, 
4d5p4f, 
4d5p5f, and
4d4f5f 
for the even parity and 
4d$^2$5p, 
4d$^2$6p, 
4d$^2$4f, 
4d$^2$5f, 
4d5s5p, 
4d5s4f, 
4d5s5f, 
4d5p5d, 
4d4f5d, and 
4d5d5f 
for the odd parity. The core-polarization parameters were the dipole polarizability of a \ion{Mo}{vii} 
ionic core reported by \citet{fragaetal1976}, i.e., $\alpha_\mathrm{d} = 1.82$\,a.u., and the cut-off radius 
corresponding to the HFR mean value $\left<r\right>$ of the outermost core orbital (4p), i.e., 
$r_\mathrm{c} = 1.20$\,a.u. Using the experimental energy levels reported by \citet{sugarmusgrove1988} and 
\citet{cabezaetal1989}, the radial integrals (average energy, Slater, spin-orbit and effective interaction parameters) of 
4d$^3$, 
4d$^2$5s,
4d$^2$6s, 
4d$^2$5d, and 
4d$^2$5p configurations were adjusted by a well-established least-squares fitting process that minimizes the differences 
between computed and experimental energies.
 
For \ion{Mo}{v}, the configurations retained in the HFR model were 
4d$^2$, 
4d5s, 
4d6s, 
4d5d, 
4d6d, 
4d5g, 
5s$^2$, 
5p$^2$, 
5d$^2$, 
4f$^2$, 
5s5d, 
5s6s, 
5p4f, 
5p5f, 
4p$^5$4d$^2$4f, 
4p$^5$4d$^2$5f, and
4p$^5$4d$^2$5p 
for the even parity and 
4d5p, 
4d6p, 
4d4f, 
4d5f, 
4d6f, 
4d7f, 
4d8f, 
4d9f, 
5s5p, 
5p5d, 
5s4f, 
5s5f, 
4f5d, 
4p$^5$4d$^3$, 
4p$^5$4d$^2$5s, and
4p$^5$4d$^2$5d 
for the odd parity. In this ion, the semi-empirical process was performed to optimize the radial 
integrals corresponding to 
4d$^2$, 
4d5s, 
4d6s, 
4d5d, 
4d5g, 
5s$^2$, 
5p$^2$, 
5s5d, 
5s6s, 
4d5p, 
4d6p, 
4d4f, 
4d5f, 
4d6f, 
4d7f, 
4d8f, 
4d9f, 
5s5p, 
5s4f, and 
4p$^5$4d$^3$ 
configurations, which use the experimental energy levels reported by \citet{readertauheed2015}. 
Core-polarization effects were estimated using a dipole polarizability value corresponding to a 
\ion{Mo}{viii} ionic core taken from \citet{fragaetal1976}, i.e.,  $\alpha_\mathrm{d} = 1.48$\,a.u., and a cut-off radius 
equal to 1.20\,a.u.

In the case of \ion{Mo}{vi}, the 
4d, 
5d, 
6d, 
7d, 
8d, 
5s, 
6s, 
7s, 
8s, 
5g, 
6g, 
7g, 
8g, 
7i, 
8i, 
4p$^5$4d5p, 
4p$^5$4d4f, and
4p$^5$4d5f 
even configurations and the 
5p, 
6p, 
7p, 
8p, 
9p, 
10p, 
11p, 
4f, 
5f, 
6f, 
7f, 
8f, 
9f, 
6h, 
7h, 
8h, 
8k, 
4p$^5$4d$^2$, 
4p$^5$4d5s, and
4p$^5$4d5d 
odd configurations were explicitly included in the HFR model with the same core-polarization parameters as those 
considered for \ion{Mo}{v}. The semi-empirical optimization process was carried out to fit the radial parameters 
in the 
nd (n = 4 – 8), 
ns (n = 5 – 8), 
ng (n = 5 – 8), 
ni (n = 7 – 8), 
np (n = 5 – 11), 
nf (n = 4 – 9), 
nh (n = 6 – 8), 
8k, 
4p$^5$4d$^2$, and 
4p$^5$4d5s configurations
using the experimental energy levels published by \citet{reader2010}.

Finally, for \ion{Mo}{vii}, the HFR multiconfiguration expansions included the 
4p$^6$, 
4p$^5$5p, 
4p$^5$6p, 
4p$^5$4f, 
4p$^5$5f, 
4p$^5$6f, 
4s4p$^6$4d, 
4s4p$^6$5d, 
4s4p$^6$6d, 
4s4p$^6$5s, 
4s4p$^6$6s, 
4p$^4$4d$^2$, 
4p$^4$4d5s, and
4p$^4$5s$^2$ 
even configurations and the 
4p$^5$4d, 
4p$^5$5d, 
4p$^5$6d, 
4p$^5$5s, 
4p$^5$6s, 
4p$^5$7s, 
4p$^5$8s, 
4p$^5$9s, 
4p$^5$10s, 
4p$^5$5g, 
4p$^5$6g, 
4s4p$^6$5p, 
4s4p$^6$6p, 
4s4p$^6$4f, 
4s4p$^6$5f, 
4s4p$^6$6f, 
4p$^4$4d5p, and
4p$^4$4d4f 
odd configurations. Here, because some configurations with open 4s and 4p orbitals were explicitly 
included in the physical model, the core-polarization effects were estimated by considering a \ion{Mo}{xv} ionic core 
with the corresponding dipole polarizability value taken from \citet{johnsonetal1983}, i.e., $\alpha_\mathrm{d} = 0.058$\,a.u., 
and a cut-off radius equal to 0.41 a.u, which corresponds to the HFR mean value $\left<r\right>$ of the 3d subshell. 
The fitting process was then carried out with the experimental energy levels classified by \citet{sugarmusgrove1988} and 
\citet{shiraietal2000} to adjust the radial parameters that characterize the 
4p$^6$, 
4p$^5$5p, 
4p$^5$4f, 
4p$^5$5f, 
4s4p$^6$4d, 
4p$^4$4d$^2$, 
4p$^5$4d, 
4p$^5$5d, 
4p$^5$ns (n = 5 – 10), 
4p$^5$5g, and 
4s4p$^6$5p configurations.
The parameters adopted in our computations are summarized in Tables 
\ref{tab:moiv:para} - \ref{tab:movii:para} and
computed and available experimental energies are compared in Tables 
\ref{tab:moiv:ener} - \ref{tab:movii:ener}, for \ion{Mo}{iv-vii}, respectively.

\onllongtab{
\begin{longtable}{llrrrl}
\caption{\label{tab:moiv:para}Radial parameters (in cm$^{-1}$) adopted for the calculations in \ion{Mo}{iv}.} \\
\hline\hline
\noalign{\smallskip}
Configuration & Parameter & HFR & Fitted & Ratio & Note$^a$ \\ 
\noalign{\smallskip}
\hline
\endfirsthead
\caption{continued.}\\
\hline\hline
\noalign{\smallskip}
Configuration & Parameter & HFR & Fitted & Ratio & Note$^a$ \\
\noalign{\smallskip}
\hline
\noalign{\smallskip}
\endhead
\hline
\noalign{\smallskip}
\endfoot
\noalign{\smallskip}
\multicolumn{6}{c}{Even parity} \\
4d$^3$           & E$_{av}$          &   17281 &  16028 &       &    \\
                 & F$^2$(4d,4d)      &   63593 &  53099 & 0.835 &    \\
                 & F$^4$(4d,4d)      &   41828 &  36184 & 0.865 &    \\
                 & $\alpha$          &       0 &      9 &       &    \\
                 & $\beta$           &       0 & $-$491 &       &    \\
                 & $\zeta$$_{4d}$    &     812 &    810 & 0.997 &    \\
4d$^2$5s         & E$_{av}$          &   75210 &  73187 &       &    \\
\hline
\end{longtable}
$^{(a)}$ F: parameter fixed in the fitting process.
}

\onllongtab{
\begin{longtable}{llrrrl}
\caption{\label{tab:mov:para}Radial parameters (in cm$^{-1}$) adopted for the calculations in \ion{Mo}{v}.} \\
\hline\hline
\noalign{\smallskip}
Configuration & Parameter & HFR & Fitted & Ratio & Note$^a$ \\ 
\noalign{\smallskip}
\hline
\endfirsthead
\caption{continued.}\\
\hline\hline
\noalign{\smallskip}
Configuration & Parameter & HFR & Fitted & Ratio & Note$^a$ \\
\noalign{\smallskip}
\hline
\noalign{\smallskip}
\endhead
\hline
\noalign{\smallskip}
\endfoot
\noalign{\smallskip}
\multicolumn{6}{c}{Even parity} \\
4d$^2$           & E$_{av}$          &   16700 &  15960 &       &    \\
                 & F$^2$(4d,4d)      &   67562 &  55311 & 0.819 &    \\
                 & F$^4$(4d,4d)      &   44738 &  35260 & 0.788 &    \\
                 & $\alpha$          &       0 &     76 &       &    \\
                 & $\beta$           &       0 & $-$166 &       &    \\
                 & $\zeta$$_{4d}$    &     911 &    920 & 1.009 &    \\
4d5s             & E$_{av}$          &  104520 &  99547 &       &    \\
\hline
\end{longtable}
$^a$ F: Fixed parameter value ; R1, R2, R3: ratios of these parameters were fixed in the fitting process.
}

\onllongtab{
\begin{longtable}{llrrrl}
\caption{\label{tab:movi:para}Radial parameters (in cm$^{-1}$) adopted for the calculations in \ion{Mo}{vi}.} \\
\hline\hline
\noalign{\smallskip}
Configuration & Parameter & HFR & Fitted & Ratio & Note$^a$ \\ 
\noalign{\smallskip}
\hline
\endfirsthead
\caption{continued.}\\
\hline\hline
\noalign{\smallskip}
Configuration & Parameter & HFR & Fitted & Ratio & Note$^a$ \\
\noalign{\smallskip}
\hline
\noalign{\smallskip}
\endhead
\hline
\noalign{\smallskip}
\endfoot
\noalign{\smallskip}
\multicolumn{6}{c}{Even parity} \\
4d               & E$_{av}$          &    6940 &   6979 &       &    \\
                 & $\zeta$$_{4d}$     &    1014 &   1040 & 1.026 &    \\
5d               & E$_{av}$          &  289239 & 285322 &       &    \\
                 & $\zeta$$_{5d}$     &     264 &    303 & 1.148 &    \\
6d               & E$_{av}$          &  391958 & 387186 &       &    \\
                 & $\zeta$$_{6d}$     &     126 &    146 & 1.161 &    \\
7d               & E$_{av}$          &  444684 & 440130 &       &    \\
\hline
\end{longtable}
$^a$ F: Fixed parameter value ; R1, R2, R3: ratios of these parameters were fixed in the fitting process.
}

\onllongtab{
\begin{longtable}{llrrrl}
\caption{\label{tab:movii:para}Radial parameters (in cm$^{-1}$) adopted for the calculations in \ion{Mo}{vii}.} \\
\hline\hline
\noalign{\smallskip}
Configuration & Parameter & HFR & Fitted & Ratio & Note$^a$ \\ 
\noalign{\smallskip}
\hline
\endfirsthead
\caption{continued.}\\
\hline\hline
\noalign{\smallskip}
Configuration & Parameter & HFR & Fitted & Ratio & Note$^a$ \\
\noalign{\smallskip}
\hline
\noalign{\smallskip}
\endhead
\hline
\noalign{\smallskip}
\endfoot
\noalign{\smallskip}
\multicolumn{6}{c}{Even parity} \\
4p$^6$           & E$_{av}$          &       0 &  21119 &       &    \\
4p$^5$5p         & E$_{av}$          &  561281 & 562851 &       &    \\
                 & $\zeta$$_{4p}$    &   14921 &  15363 & 1.030 &    \\
                 & $\zeta$$_{5p}$    &    3488 &   4002 & 1.147 &    \\
                 & F$^2$(4p,5p)      &   29857 &  24371 & 0.816 &    \\
                 & G$^0$(4p,5p)      &    6158 &   5332 & 0.866 &    \\
                 & G$^2$(4p,5p)      &    8430 &   7935 & 0.941 &    \\
\hline
\end{longtable}
$^a$ F: Fixed parameter value.
}

\onllongtab{
\begin{longtable}{rrrrl}
\caption{\label{tab:moiv:ener} Comparison between available experimental and calculated energy levels in \ion{Mo}{iv}. Energies are given in cm$^{-1}$.} \\
\hline\hline
\noalign{\smallskip}
E$_{exp}$$^a$  & E$_{calc}$$^b$ & $\Delta$E & J & Leading components (in \%) in $LS$ coupling$^c$ \\ 
\noalign{\smallskip}
\hline
\endfirsthead
\caption{continued.}\\
\hline\hline
\noalign{\smallskip}
E$_{exp}$$^a$  & E$_{calc}$$^b$ & $\Delta$E & J & Leading components (in \%) in $LS$ coupling$^c$ \\
\noalign{\smallskip}
\hline
\noalign{\smallskip}
\endhead
\hline
\noalign{\smallskip}
\endfoot
\noalign{\smallskip}
\multicolumn{5}{c}{Even parity} \\
        0.00  &   $-$34 &     34 &  1.5  &  99  4d$^3$ $^4$F  \\  
      778.02  &     742 &     36 &  2.5  &  99  4d$^3$ $^4$F  \\  
     1759.76  &    1716 &     44 &  3.5  &  99  4d$^3$ $^4$F  \\  
     2864.58  &    2817 &     48 &  4.5  &  98  4d$^3$ $^4$F  \\  
    10330.59  &   10370 &  $-$39 &  1.5  &  82  4d$^3$ $^4$P   +  16  4d$^3$ $^2$P  \\  
    10337.77  &   10414 &  $-$76 &  0.5  &  92  4d$^3$ $^4$P   +   8  4d$^3$ $^2$P  \\
    11580.50  &   11661 &  $-$80 &  3.5  &  99  4d$^3$ $^2$G  \\
\hline
\end{longtable}
\noindent
$^{(a)}$ From  \citet{sugarmusgrove1988} and \citet{cabezaetal1989}.\\
$^{(b)}$ This work.\\
$^{(c)}$ Only the first three components that are larger than 5\% are given.
}

\onllongtab{
\begin{longtable}{rrrrl}
\caption{\label{tab:mov:ener} Comparison between available experimental and calculated energy levels in \ion{Mo}{v}. Energies are given in cm$^{-1}$.} \\
\hline\hline
\noalign{\smallskip}
E$_{exp}$$^a$  & E$_{calc}$$^b$ & $\Delta$E & J & Leading components (in \%) in $LS$ coupling$^c$ \\ 
\noalign{\smallskip}
\hline
\endfirsthead
\caption{continued.}\\
\hline\hline
\noalign{\smallskip}
E$_{exp}$$^a$  & E$_{calc}$$^b$ & $\Delta$E & J & Leading components (in \%) in $LS$ coupling$^c$ \\
\noalign{\smallskip}
\hline
\noalign{\smallskip}
\endhead
\hline
\noalign{\smallskip}
\endfoot
\noalign{\smallskip}
\multicolumn{5}{c}{Even parity} \\
        0.00  &      12 &  $-$12 &  2  &   97  4d$^2$ $^3$F \\  
     1577.17  &    1577 &      0 &  3  &   99  4d$^2$ $^3$F \\  
     3357.08  &    3345 &     12 &  4  &   98  4d$^2$ $^3$F \\   
    10190.11  &   10186 &      4 &  2  &   76  4d$^2$ $^1$D  +  21  4d$^2$ $^3$P \\  
    11161.31  &   11149 &     12 &  0  &   98  4d$^2$ $^3$P \\
    11806.93  &   11808 &   $-$1 &  1  &   99  4d$^2$ $^3$P \\  
    13408.30  &   13423 &  $-$15 &  2  &   78  4d$^2$ $^3$P  +  21  4d$^2$ $^1$D \\
\hline
\end{longtable}
\noindent
$^{(a)}$ From \citet{readertauheed2015}.\\
$^{(b)}$ This work.\\
$^{(c)}$ Only the first three components that are larger than 5\% are given.
}

\onllongtab{
\begin{longtable}{rrrrl}
\caption{\label{tab:movi:ener} Comparison between available experimental and calculated energy levels in \ion{Mo}{vi}. Energies are given in cm$^{-1}$.} \\
\hline\hline
\noalign{\smallskip}
E$_{exp}$$^a$  & E$_{calc}$$^b$ & $\Delta$E & J & Leading components (in \%) in $LS$ coupling$^c$ \\ 
\noalign{\smallskip}
\hline
\endfirsthead
\caption{continued.}\\
\hline\hline
\noalign{\smallskip}
E$_{exp}$$^a$  & E$_{calc}$$^b$ & $\Delta$E & J & Leading components (in \%) in $LS$ coupling$^c$ \\
\noalign{\smallskip}
\hline
\noalign{\smallskip}
\endhead
\hline
\noalign{\smallskip}
\endfoot
\noalign{\smallskip}
\multicolumn{5}{c}{Even parity} \\
        0.00  &       0 &      0 & 1.5  &  99  4d $^2$D  \\   
     2583.50  &    2583 &      0 & 2.5  &  99  4d $^2$D  \\   
   119725.62  &  119725 &      0 & 0.5  &  99  5s $^2$S  \\
   282825.59  &  282825 &      0 & 1.5  &  99  5d $^2$D  \\   
   283610.94  &  283611 &      0 & 2.5  &  99  5d $^2$D  \\   
   313806.81  &  313807 &      0 & 0.5  & 100  6s $^2$S  \\
   386165.59  &  386166 &      0 & 1.5  &  99  6d $^2$D  \\
\hline
\end{longtable}
\noindent
$^{(a)}$ From \citet{reader2010}.\\
$^{(b)}$ This work.\\
$^{(c)}$ Only the first three components that are larger than 5\% are given.
}

\onllongtab{
\begin{longtable}{rrrrl}
\caption{\label{tab:movii:ener} Comparison between available experimental and calculated energy levels in \ion{Mo}{vii}. Energies are given in cm$^{-1}$.} \\
\hline\hline
\noalign{\smallskip}
E$_{exp}$$^a$  & E$_{calc}$$^b$ & $\Delta$E & J & Leading components (in \%) in $LS$ coupling$^c$ \\ 
\noalign{\smallskip}
\hline
\endfirsthead
\caption{continued.}\\
\hline\hline
\noalign{\smallskip}
E$_{exp}$$^a$  & E$_{calc}$$^b$ & $\Delta$E & J & Leading components (in \%) in $LS$ coupling$^c$ \\
\noalign{\smallskip}
\hline
\noalign{\smallskip}
\endhead
\hline
\noalign{\smallskip}
\endfoot
\noalign{\smallskip}
\multicolumn{5}{c}{Even parity} \\
        0.00  &       0 &      0 & 0 &   97  4p$^6$ $^1$S \\
   542265.29  &  542456 & $-$191 & 1 &   80  4p$^5$5p $^3$S  +   17  4p$^5$5p $^3$P \\ 
   544663.10  &  544655 &      8 & 1 &   73  4s4p$^6$4d $^3$D  +   12  4p$^4$($^1$D)4d$^2$ $^3$D  +    8  4p$^5$4f $^3$D \\ 
   545906.83  &  545883 &     24 & 2 &   73  4s4p$^6$4d $^3$D  +   11  4p$^4$($^1$D)4d$^2$ $^3$D  +    8  4p$^5$4f $^3$D \\ 
   548077.57  &  547987 &     91 & 2 &   55  4p$^5$5p $^3$D  +   35  4p$^5$5p $^1$D  +    8  4p$^5$5p $^3$P \\ 
   548120.43  &  548120 &      0 & 3 &   74  4s4p$^6$4d  $^3$D  +   12  4p$^4$($^1$D)4d$^2$ $^3$D  +    7  4p$^5$4f $^3$D \\ 
   552329.97  &  552000 &    330 & 3 &   98  4p$^5$5p $^3$D \\
\hline
\end{longtable}
\noindent
$^{(a)}$ From \citet{sugarmusgrove1988} and \citet{shiraietal2000}.\\
$^{(b)}$ This work.\\
$^{(c)}$ Only the first three components that are larger than 5\% are given.
}

Tables 11 - 14 give the weighted HFR oscillator strengths 
($\log gf$) and transition probabilities ($gA$, in s$^{-1}$) for \ion{Mo}{iv-vii}, respectively, and  
the numerical values (in cm$^{-1}$) of lower and upper energy levels and the corresponding wavelengths (in \AA). 
In the last column of each table, we also give the absolute value of the cancellation factor CF, as defined by \citet{cowan1981}. 
We note that very low values of this  factor (typically $< 0.05$) indicate strong cancellation effects in the
calculation of line strengths. In these cases, the corresponding $gf$ and $gA$ values could be very inaccurate 
and, therefore, need to be considered with some care. However, very few of the transitions that appear in 
Tables 11 - 14 are affected. 
These tables are provided via the newly developed GAVO T\"ubingen Oscillator Strengths Service 
TOSS\footnote{\url{http://dc.g-vo.org/TOSS}} that is briefly described in Sect.\,\ref{sect:toss}.





\begin{figure}
   \resizebox{\hsize}{!}{\includegraphics{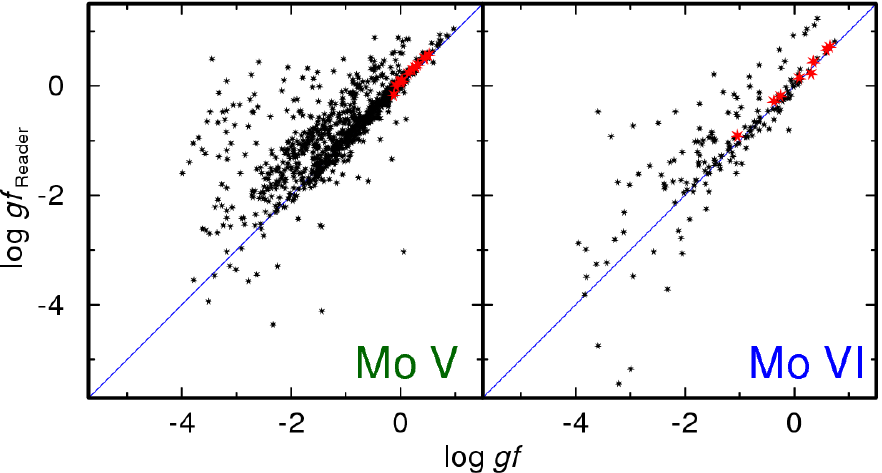}}
    \caption{Comparison of our weighted oscillator strengths to those of
             \citet{readertauheed2015} for \ion{Mo}{v} (left panel) and of
             \citet{reader2010} for \ion{Mo}{vi} (right).
             The larger, red symbols refer to the lines identified in \re (Table\,\ref{tab:fcompare}).
            }
   \label{fig:fcomp}
\end{figure}

\begin{figure}
   \resizebox{\hsize}{!}{\includegraphics{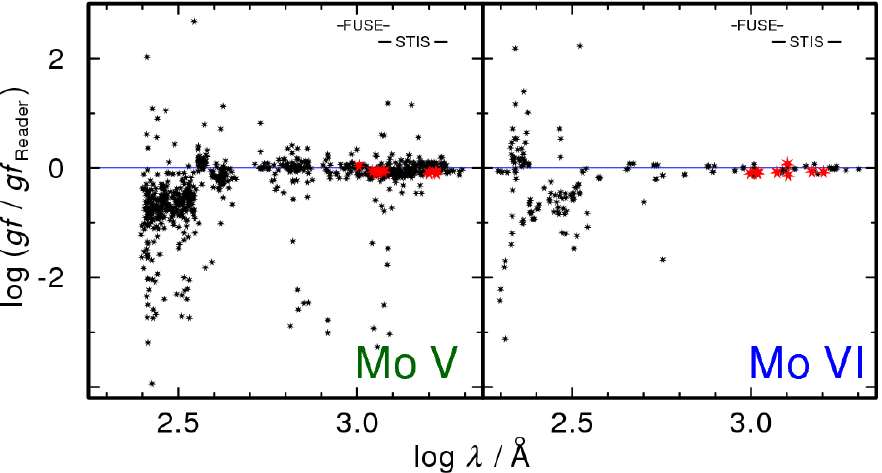}}
    \caption{Ratio of our weighted oscillator strengths and those of
             \citet{readertauheed2015} for \ion{Mo}{v} (left panel) and of
             \citet{reader2010} for \ion{Mo}{vi} (right). The wavelength ranges
             of our FUSE and HST/STIS spectra are marked.
             The larger, red symbols refer to the lines identified in \re (Table\,\ref{tab:fcompare}).
            }
   \label{fig:wfcomp}
\end{figure}

Oscillator strengths were published by 
\citet{readertauheed2015} for 923 lines of \ion{Mo}{v} and by
\citet{reader2010} for 245 lines of \ion{Mo}{vi}.
We compared $\log gf$ values and wavelengths of those lines whose positions agree with those of our lines
within $\Delta\lambda \le 0.02\,\mathrm{\AA}$ (these are 921 lines of \ion{Mo}{v}, 178 lines of \ion{Mo}{vi})
(Figs.\ref{fig:fcomp} and \ref{fig:wfcomp}).
In general, we find a rather good agreement, although our new $\log gf$-values seem to be, on average, smaller than those 
previously published. This can be explained by the fact that our calculations explicitly include a larger set of 
interacting configurations, in particular with an open 4p subshell, as well as a pseudo-potential modeling of the
remaining core-valence electronic correlations.
For the \ion{Mo}{v} and \ion{Mo}{vi} lines that were identified in \re (Table\,\ref{tab:fcompare}, Fig.\,\ref{fig:molines})
and were used for the abundance determination, the $\log gf$-values are almost identical.

\onltab{
\begin{table*}\centering
\caption{Comparison of wavelengths and $\log gf$ values of the identified \ion{Mo}{v} and \ion{Mo}{vi} lines to literature values.}         
\label{tab:fcompare}
\begin{tabular}{lr@{\,\,--\,\,}lrrrr@{.}lr@{.}lcr@{.}lr@{.}l}
\hline
\hline
\noalign{\smallskip}
\multicolumn{3}{c}{}              & \multicolumn{2}{c}{Energy\,/\,cm$^{-1}$} & \ &
                                      \multicolumn{2}{c}{$\lambda$\,/\,\AA} & \multicolumn{2}{c}{$\log gf$} & \ &
                                      \multicolumn{2}{c}{$\lambda$\,/\,\AA} & \multicolumn{2}{c}{$\log gf$} \\
\noalign{\smallskip}
\cline{4-5}
\cline{7-10}
\cline{12-15}
\multicolumn{14}{c}{} \vspace{-6mm}\\              
Ion &\multicolumn{2}{c}{Transition} \vspace{-2mm}\\
\noalign{\smallskip}
\multicolumn{3}{c}{}             & Lower level & Upper level & &
                                   \multicolumn{4}{c}{this work} & &
                                   \multicolumn{4}{c}{literature} \\
\noalign{\smallskip}
\hline
\noalign{\smallskip}
\ion{Mo}{v}  & 5p     $^3$P$^\mathrm{o}_{2.0}$ & 6s $^5$5/2[5/2]$_{3.0}$     & 157851\hspace{2mm}\hbox{} & 256676\hspace{2mm}\hbox{} & & 1011&889 & $-0$&$12$ && 1011&891 & $-0$&$16$\tablefootmark{a} \\    
             & 5p     $^1$D$^\mathrm{o}_{2.0}$ & 5d $^3$F$_{2.0}$            & 146977\hspace{2mm}\hbox{} & 237760\hspace{2mm}\hbox{} & & 1101&530 & $ 0$&$01$ && 1101&529 & $ 0$&$07$\tablefootmark{a} \\    
             & 5p     $^3$D$^\mathrm{o}_{1.0}$ & 5d $^3$F$_{2.0}$            & 148949\hspace{2mm}\hbox{} & 237760\hspace{2mm}\hbox{} & & 1125&988 & $ 0$&$22$ && 1125&990 & $ 0$&$31$\tablefootmark{a} \\    
             & 5p     $^3$F$^\mathrm{o}_{3.0}$ & 5d $^3$F$_{3.0}$            & 151195\hspace{2mm}\hbox{} & 239069\hspace{2mm}\hbox{} & & 1137&995 & $ 0$&$32$ && 1137&999 & $ 0$&$39$\tablefootmark{a} \\    
             & 5p     $^3$D$^\mathrm{o}_{3.0}$ & 5d $^3$F$_{4.0}$            & 153040\hspace{2mm}\hbox{} & 240110\hspace{2mm}\hbox{} & & 1148&502 & $ 0$&$45$ && 1148&504 & $ 0$&$52$\tablefootmark{a} \\    
             & 5p     $^3$F$^\mathrm{o}_{3.0}$ & 5d $^3$G$_{4.0}$            & 151195\hspace{2mm}\hbox{} & 235496\hspace{2mm}\hbox{} & & 1186&227 & $ 0$&$52$ && 1186&230 & $ 0$&$57$\tablefootmark{a} \\    
             & 5p     $^3$D$^\mathrm{o}_{1.0}$ & 5d $^3$D$_{1.0}$            & 148949\hspace{2mm}\hbox{} & 233190\hspace{2mm}\hbox{} & & 1187&061 & $ 0$&$03$ && 1187&061 & $ 0$&$10$\tablefootmark{a} \\    
             & 5s     $^3$D$_{3.0}$           & 5p $^3$P$^\mathrm{o}_{2.0}$  &  94835\hspace{2mm}\hbox{} & 157851\hspace{2mm}\hbox{} & & 1586&898 & $ 0$&$01$ && 1586&890 & $ 0$&$09$\tablefootmark{a} \\    
             & 5s     $^1$D$_{2.0}$           & 5p $^1$P$^\mathrm{o}_{1.0}$  &  99380\hspace{2mm}\hbox{} & 162257\hspace{2mm}\hbox{} & & 1590&414 & $-0$&$08$ && 1590&415 & $ 0$&$01$\tablefootmark{a} \\    
             & 5s     $^1$D$_{2.0}$           & 5p $^1$F$^\mathrm{o}_{3.0}$  &  99380\hspace{2mm}\hbox{} & 159857\hspace{2mm}\hbox{} & & 1653&541 & $ 0$&$28$ && 1653&541 & $ 0$&$35$\tablefootmark{a} \\    
             & 5s     $^3$D$_{3.0}$           & 5p $^3$F$^\mathrm{o}_{4.0}$  &  94835\hspace{2mm}\hbox{} & 155032\hspace{2mm}\hbox{} & & 1661&215 & $ 0$&$44$ && 1661&215 & $ 0$&$51$\tablefootmark{a} \\    
             & 5s     $^3$D$_{2.0}$           & 5p $^3$D$^\mathrm{o}_{3.0}$  &  93111\hspace{2mm}\hbox{} & 153040\hspace{2mm}\hbox{} & & 1668&662 & $ 0$&$15$ && 1668&660 & $ 0$&$24$\tablefootmark{a} \\    
\noalign{\smallskip}                                                                                             
\ion{Mo}{vi} & 5p     $^2$P$^\mathrm{o}_{1/2}$ & 5d $^2$D$_{3/2}$            & 182404\hspace{2mm}\hbox{} & 282826\hspace{2mm}\hbox{} & &  995&806 & $ 0$&$35$ &&  995&811 & $ 0$&$44$\tablefootmark{b} \\    
             & 5p     $^2$P$^\mathrm{o}_{3/2}$ & 5d $^2$D$_{5/2}$            & 187331\hspace{2mm}\hbox{} & 283611\hspace{2mm}\hbox{} & & 1038&640 & $ 0$&$59$ && 1038&642 & $ 0$&$67$\tablefootmark{b} \\    
             & 5p     $^2$P$^\mathrm{o}_{3/2}$ & 5d $^2$D$_{3/2}$            & 187331\hspace{2mm}\hbox{} & 282826\hspace{2mm}\hbox{} & & 1047&182 & $-0$&$37$ && 1047&184 & $-0$&$28$\tablefootmark{b} \\    
             & 5d     $^2$D$_{5/2}$           & 5f $^2$F$^\mathrm{o}_{7/2}$  & 283611\hspace{2mm}\hbox{} & 368203\hspace{2mm}\hbox{} & & 1182&142 & $ 0$&$65$ && 1182&143 & $ 0$&$72$\tablefootmark{b} \\    
             & 5g     $^2$G$_{9/2}$           & 7h $^2$H$^\mathrm{o}_{11/2}$ & 395184\hspace{2mm}\hbox{} & 474297\hspace{2mm}\hbox{} & & 1264&023 & $ 0$&$31$ && 1264&052 & $ 0$&$32$\tablefootmark{b,c} \\    
             & 4d$^2$ $^2$F$^\mathrm{o}_{7/2}$ & 5g $^2$G$_{7/2}$            & 316477\hspace{2mm}\hbox{} & 395184\hspace{2mm}\hbox{} & & 1270&523 & $-1$&$04$ && 1270&520 & $-0$&$91$\tablefootmark{b} \\    
             & 4d$^2$ $^2$F$^\mathrm{o}_{7/2}$ & 6d $^2$$D_{5/2}$            & 316477\hspace{2mm}\hbox{} & 386552\hspace{2mm}\hbox{} & & 1427&030 & $-2$&$22$ && \multicolumn{2}{c}{---} & 
                                                                                                                                                               \multicolumn{2}{c}{---} \\
             & 5s     $^2$S$_{1/2}$           & 5p $^2$P$^\mathrm{o}_{3/2}$  & 119726\hspace{2mm}\hbox{} & 187331\hspace{2mm}\hbox{} & & 1479&168 & $ 0$&$09$ && 1479&168 & $ 0$&$15$\tablefootmark{b} \\    
             & 5s     $^2$$S_{1/2}$           & 5p $^2$P$^\mathrm{o}_{1/2}$  & 119726\hspace{2mm}\hbox{} & 182404\hspace{2mm}\hbox{} & & 1595&435 & $-0$&$25$ && 1595&435 & $-0$&$18$\tablefootmark{b} \\    
\noalign{\smallskip}
\hline
\end{tabular}
\tablefoot{
\tablefoottext{a}{\citet{readertauheed2015},}
\tablefoottext{b}{\citet{reader2010}, \ionw{Mo}{vi}{1427.030} is not listed,}
\tablefoottext{c}{blend with  5g $^2$G$_{7/2}$\,\,--\,\,7h $^2$H$^\mathrm{o}_{9/2}$, $\lambda$\,1264.017\,\AA, $\log gf = 0.23$.}
}
\end{table*}  
}

\section{The GAVO service TOSS}
\label{sect:toss}

In the framework of the GAVO project, we  developed the new, registered VO service TOSS.
It is designed to provide easy access to calculated oscillator strengths and transition probabilities 
of any kind in VO-compliant format. 

Line data is stored in terms of the Spectral Line Data Model
\citep{std:ssldm} and is accessible through a web
browser interface, via the Simple Line
Access Protocol SLAP \citep{std:slap}, and via the Table Access Protocol
TAP \citep{std:TAP}.
The browser-based interface offers a web form, which allows conventional
queries by wavelength, element, ionisation stage, etc., exporting to
various tabular formats and also directly into user programs via
SAMP\footnote{The Simple Application Messaging Protocol is a VO-defined
standard protocol facilitating seamless and fast data exchange on user
desktops.}.

The SLAP interface can be used from specialized programs (``SLAP
clients'') like VOSpec \citep{2005ASPC..347..198O}; this is normally 
transparent to the user; in a server selector, the service will
typically appear under its short name TOSS SLAP.

The TAP interface allows database queries against the line database,
including comparisons with user-provided data (``uploads'').  In the TAP
dialogs of applications like TOPCAT, the user should look for ``GAVO DC'' or manually
enter the access URL \url{http://dc.g-vo.org/tap}.  As the discussion of TAP's
possibilities is beyond the scope of this paper, see on-line examples for TOSS service
usage\footnote{\url{http://dc.g-vo.org/toss/q/q/examples}}.

\section{Photospheric abundances}
\label{sect:abundances}

To improve the simulation of the background opacity, 
we included As in our calculations for \gb and \re and determined its abundance
from these models (Sect.\,\ref{sect:as}). Then, we included Sn for \re (Sect.\,\ref{sect:snre})
because our new HST/STIS observation provides access to \ion{Sn}{iv} lines that are necessary
for an abundance analysis. Based on these extended models, we considered Mo and determined 
its abundances for \re (Sect.\,\ref{sect:more}) and \gb (Sect.\,\ref{sect:mogb}).

\subsection{\gb and \re: Arsenic ($Z = 33$)}
\label{sect:as}

In the FUSE observation of \re, \citet{werneretal2012} discovered 
\ionww{As}{v}{987.65, 1029.48, 1051.6, 1056.7}.
For \ion{As}{v}, lifetimes were measured with the beam-foil technique \citep{pinningtonetal1981}.
The multiplet f-value ($0.78\pm 0.06$) of the \ion{As}{v} resonance transition was determined with the help
of arbitrarily normalized decay curve (ANDC) analyses, which were confirmed by calculations of
\citet{fischer1977},
\citet{migdalekbaylis1979}, and 
\citet{curtistheodosiou1989}. 
\citet{morton2000} lists both components in his compilation,
\ionw{}{}{987.65} (4s\,$^2$S$_\mathrm{1/2}$ -- 4p\,$^2$P$^\mathrm{o}_\mathrm{3/2}$, $f = 0.528$) and
\ionw{}{}{1029.48} (4s\,$^2$S$_\mathrm{1/2}$ -- 4p\,$^2$P$^\mathrm{o}_\mathrm{1/2}$, $f = 0.253$).
These were used by \citet{chayeretal2015} to determine arsenic mass fractions. They found
$6.3\,\times\,10^{-8}$ (6 times solar) and $1.6\,\times\,10^{-5}$ (1450 times solar) in \gb and \re, respectively.

\ionww{As}{v}{1051.6, 1056.7} belong to the 4d\,$^2$D - 4f\,$^2$F$^\mathrm{o}$ multiplet
($\lambda\lambda 1050.67, 1051.64, 1055.60, 1056.58$) but no oscillator strengths have been calculated so far.

\begin{figure}
   \resizebox{\hsize}{!}{\includegraphics{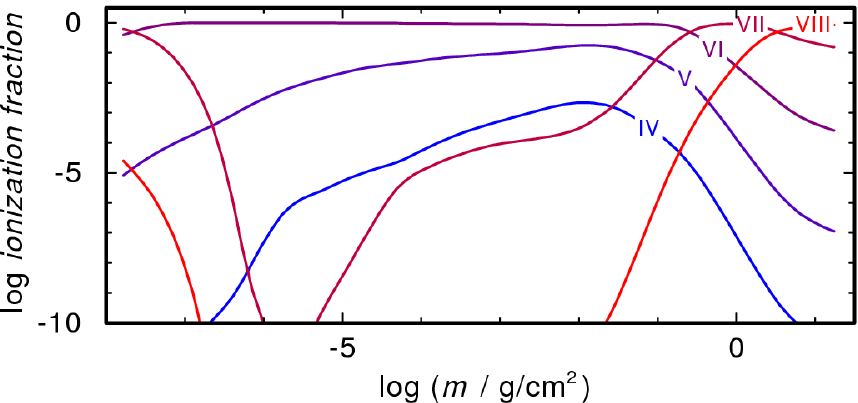}}
    \caption{Arsenic ionization fractions in our \re model.
             $m$ is the column mass, measured from the outer boundary of the model atmosphere.
            }
   \label{fig:ion_as_re}
\end{figure}

\begin{figure}
   \resizebox{\hsize}{!}{\includegraphics{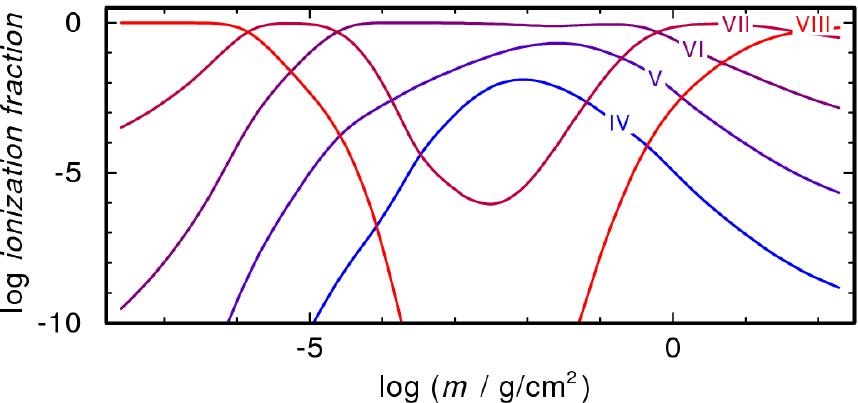}}
    \caption{As in Fig.\,\ref{fig:ion_as_re}, for \gb.
            }
   \label{fig:ion_as_gb}
\end{figure}

\begin{figure}
   \resizebox{\hsize}{!}{\includegraphics{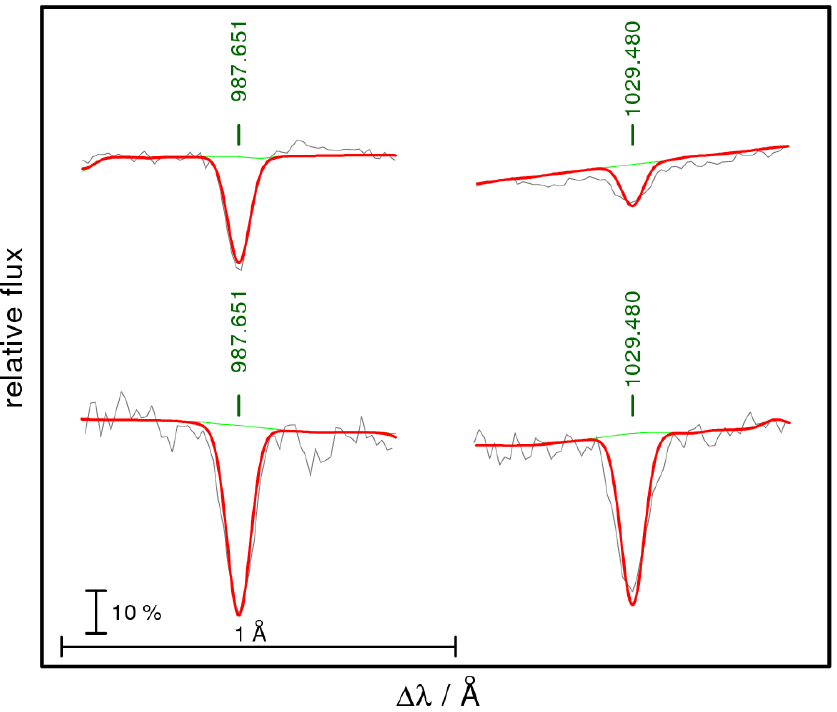}}
    \caption{Sections of our FUSE observations of 
             \gb (top row) and 
             \re (bottom row) around 
             \ionw{As}{v}{987.65} (left column) and
             \ionw{As}{v}{1029.48} (right column)
             The thick red and thin green lines show a comparison with
             theoretical spectra of two models, with and without As, respectively.
            }
   \label{fig:as}
\end{figure}

\begin{figure}
   \resizebox{\hsize}{!}{\includegraphics{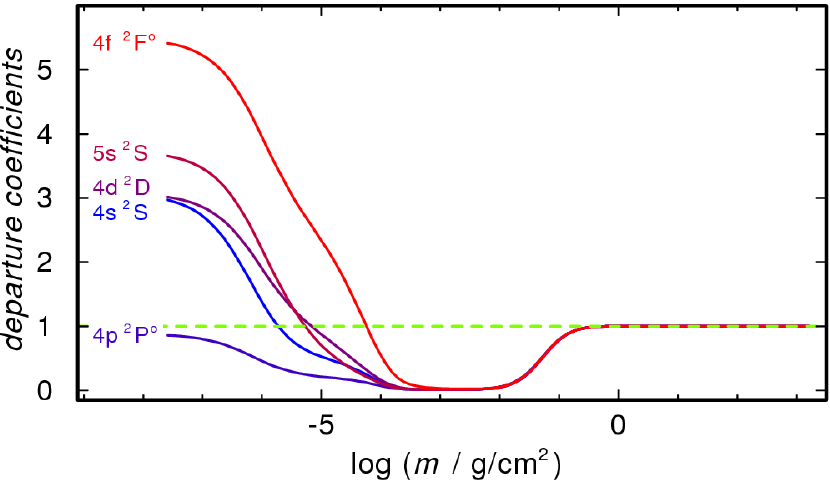}}
    \caption{Departure coefficients of the five lowest \ion{As}{v} levels in the model for \gb.
            }
   \label{fig:dep_as_gb}
\end{figure}

\begin{figure}
   \resizebox{\hsize}{!}{\includegraphics{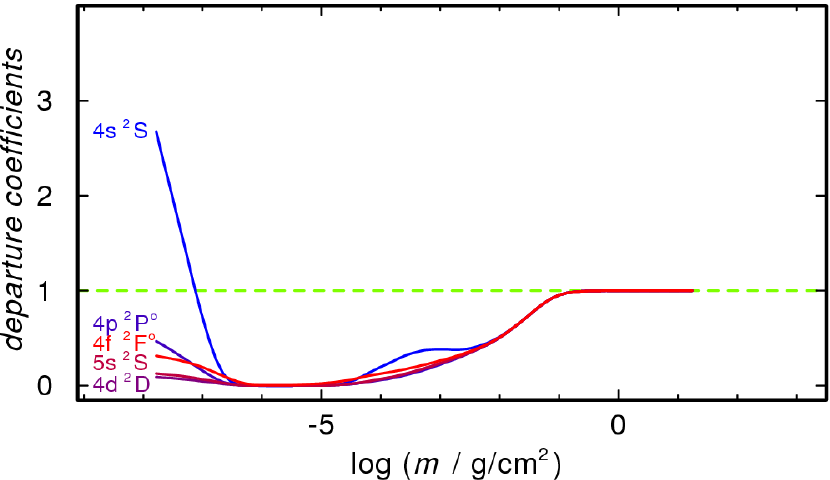}}
    \caption{As for Fig.\,\ref{fig:dep_as_gb}, for \re.
            }
   \label{fig:dep_as_re}
\end{figure}

We included the TMAD \ion{As}{iv-viii} model atom in our model-atmosphere calculations.
\ion{As}{vi} is the dominant ion in the line-forming region in both stars (Figs.\,\ref{fig:ion_as_re} and \ref{fig:ion_as_gb}). 
We reproduced  the observed \ionww{As}{v}{987.65, 1029.48} line profiles well in the FUSE spectra of \gb and \re at
mass fractions of
$3.7\,\times\,10^{-7}$ (29 times solar) and 
$8.3\,\times\,10^{-6}$ (760 times solar), 
respectively (Fig.\,\ref{fig:as}).
Our abundances deviate by factors of 6 and 0.5 for \gb and \re, respectively, from those of 
\cite{chayeretal2015}, 
who made an LTE assumption because of lack of atomic data. Both stars are out of the validity domain for
LTE modeling \citep[e.g.,][]{rauch2012}, which is corroborated by an inspection of the departure coefficients 
(ratio of NLTE to LTE occupation numbers) of the five lowest \ion{As}{v} levels in our models for \gb and \re 
that deviate from unity (Fig.\,\ref{fig:dep_as_gb} and\,\ref{fig:dep_as_re}, respectively). Therefore, 
the results of an LTE analysis may be afflicted by  a large systematic error.

\begin{figure}
   \resizebox{\hsize}{!}{\includegraphics{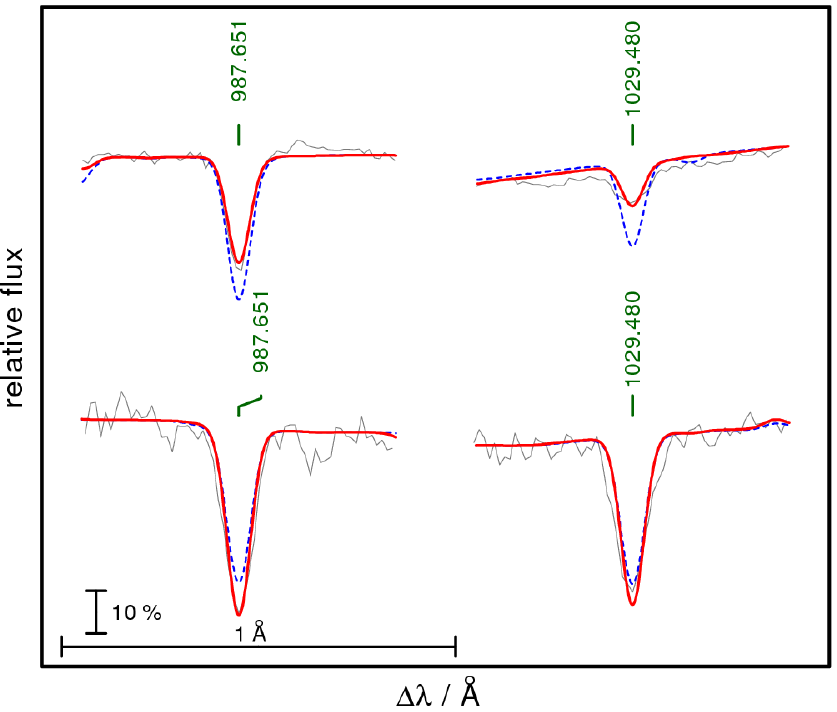}}
    \caption{As for Fig.\,\ref{fig:as},
             the thick red and dashed blue lines show a comparison with
             theoretical spectra of NLTE and LTE models, respectively.
            }
   \label{fig:as_LTE}
\end{figure}

Figure\,\ref{fig:as_LTE} shows a comparison of theoretical profiles of \ionww{As}{v}{987.65, 1029.48} that
are calculated from our NLTE model and an LTE model (calculated by TMAD based on the temperature
and density stratification of the NLTE model, LTE occupation numbers of the atomic levels are
enforced by an artificial increase of all collisional rates by a factor of $10^{20}$). 
In the case of \gb, the LTE profiles are too strong and, thus, the abundance had to be reduced 
to $1.0 \times 10^{-7}$ to match
the observation. For \re, they are too weak and, thus, a higher abundance 
of $1.7 \times 10^{-5}$
is needed to reproduce the
observation. This may explain, in part, the deviation of our As abundances from those of 
\citet{chayeretal2015}.

However, a precise abundance analysis by means of NLTE model-atmosphere techniques requires
reliable oscillator strengths, not only for the lines employed in the abundance determination, but for 
the complete model atom that is considered in the model-atmosphere and spectral-energy-distribution
calculations. Once these oscillator strengths become available for a large number of 
\ion{As}{v-vii} lines, it will be possible to construct much more complete As model ions and
a new determination of the As abundance will be more precise.

\subsection{\re: Tin ($Z = 50$)}
\label{sect:snre}

\begin{figure}
   \resizebox{\hsize}{!}{\includegraphics{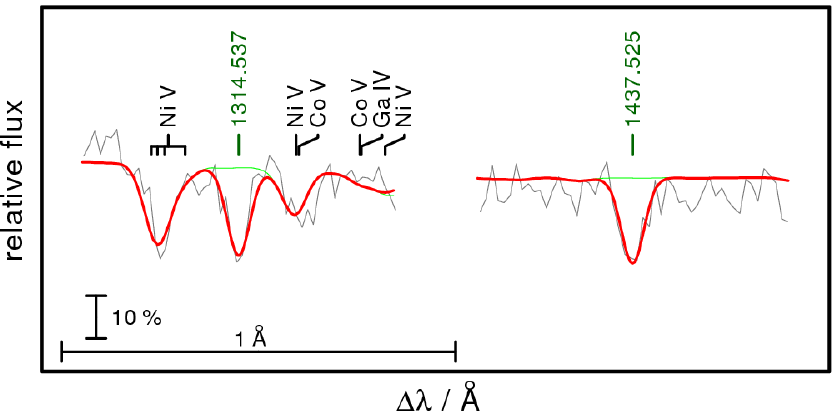}}
    \caption{Sections of our HST/STIS observations of \re around 
             \ionw{Sn}{iv}{1314.537} (left) and
             \ionw{Sn}{iv}{1427.525} (right).
             The thick, red and thin, green lines show a comparison with
             theoretical spectra of two models with and without Sn, respectively.
            }
   \label{fig:sn}
\end{figure}

\citet{rauchetal2013} measured the Sn abundance in \gb ($3.53\times 10^{-7}$, 37 times solar).
They used the lines of the resonance doublet, namely,
\ionw{Sn}{iv}{1314.537} (5s\,$^2$S$_\mathrm{1/2}$ -- 5p\,$^2$P$^\mathrm{o}_\mathrm{3/2}$, $f = 0.637$)
and
    \ionw{}{}{1437.525} (5s\,$^2$S$_\mathrm{1/2}$ -- 5p\,$^2$P$^\mathrm{o}_\mathrm{1/2}$, $f = 0.240$).
Their f-values were calculated by \citet{morton2000} based on energy levels provided by \citet{moore1958}.
These lines are visible in our HST/STIS spectrum of \re as well. To determine the Sn abundance,
we used the same model atom like \citet{rauchetal2013} and reproduced  the observed line
profiles of both lines (Fig.\,\ref{fig:sn}) well at a mass fraction of $2.04 \times 10^{-4}$ (about 22\,500 times solar).
This Sn abundance analysis will also be improved, 
as soon as reliable transition probabilities for \ion{Sn}{iv-vi} are published.

\subsection{\re: Molybdenum}
\label{sect:more}

\begin{figure}
   \resizebox{\hsize}{!}{\includegraphics{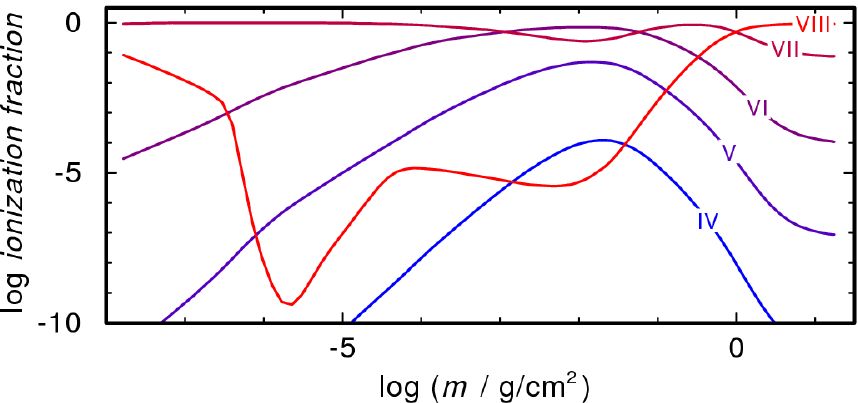}}
    \caption{Molybdenum ionization fractions in our \re model.
            }
   \label{fig:ion_re}
\end{figure}

Our \re model (\Teffw{70\,000}, $\log\,g = 7.5$) includes opacities of
H, He, C, N, O, Si, P, S, Ca, Sc, Ti, V, Cr, Mn, Fe, Co, Ni, Zn, Ga, Ge, As, Kr, Mo, Xe, and Ba.
Figure\,\ref{fig:ion_re} shows
the Mo ionization fractions in this model. \ion{Mo}{vi+vii} are the dominating ionization stages in the line-forming region ($-4.0\,\sla\,\log m\,\sla\,0.5$).
The element abundances are given in Table\,\ref{tab:abre}. 
In general, their uncertainty is about 0.2\,dex. This includes the
error propagation due to the error ranges of 
\Teff \citep[cf., e.g.,][]{venneslanz2001}, 
\logg, 
and the background opacity.

\setcounter{table}{14}

\begin{table}\centering 
  \caption{Photospheric abundances of \re.  
           IG is a generic model atom \citep{rauchdeetjen2003} comprising Ca, Sc, Ti, V, Cr, Mn, and Co.
           [X] denotes log (fraction\,/\,solar fraction) of species X.}
\label{tab:abre}
\setlength{\tabcolsep}{.4em}
\begin{tabular}{lr@{.}lr@{.}lr@{.}l}
\hline
\hline
\noalign{\smallskip}                                                                                          
                         & \multicolumn{2}{c}{Mass}   & \multicolumn{2}{c}{Number}  & \multicolumn{2}{c}{}                      \\
\cline{2-5}                     
\multicolumn{7}{c}{}                                                                                                 \vspace{-5mm}\\
Element                  & \multicolumn{2}{c}{}       & \multicolumn{2}{c}{}        & \multicolumn{2}{c}{~~~~~[X]} \vspace{-2mm}\\
                         & \multicolumn{4}{c}{Fraction}                             & \multicolumn{2}{c}{}                      \\
\cline{1-7}                     
\noalign{\smallskip}                                                                                   
\mmspr He                      & $ 9$&$74\times 10^{-1}$ & $ 9$&$92\times 10^{-1}$ & $  0$&$592$ \\
\mmspr C                       & $ 2$&$23\times 10^{-2}$ & $ 7$&$56\times 10^{-3}$ & $  0$&$974$ \\
\mmspr N                       & $ 1$&$73\times 10^{-4}$ & $ 5$&$04\times 10^{-5}$ & $ -0$&$602$ \\
\mmspr O                       & $ 1$&$97\times 10^{-3}$ & $ 5$&$01\times 10^{-4}$ & $ -0$&$464$ \\
\mmspr Si                      & $ 1$&$61\times 10^{-4}$ & $ 2$&$33\times 10^{-5}$ & $ -0$&$617$ \\
\mmspr P                       & $ 1$&$15\times 10^{-6}$ & $ 1$&$51\times 10^{-7}$ & $ -0$&$705$ \\
\mmspr S                       & $ 3$&$96\times 10^{-5}$ & $ 5$&$04\times 10^{-6}$ & $ -0$&$892$ \\
\mmspr IG                      & $ 1$&$00\times 10^{-6}$ & $ 9$&$19\times 10^{-8}$ & $ -1$&$787$ \\
\mmspr Fe                      & $<1$&$30\times 10^{-5}$ & $<9$&$50\times 10^{-7}$ & $<-1$&$967$ \\
\mmspr Ni                      & $ 7$&$26\times 10^{-5}$ & $ 5$&$04\times 10^{-6}$ & $  0$&$028$ \\
\mmspr Zn                      & $ 1$&$13\times 10^{-4}$ & $ 7$&$05\times 10^{-6}$ & $  1$&$814$ \\
\mmspr Ga                      & $ 3$&$45\times 10^{-5}$ & $ 2$&$02\times 10^{-6}$ & $  2$&$810$ \\
\mmspr Ge                      & $ 1$&$59\times 10^{-4}$ & $ 8$&$90\times 10^{-6}$ & $  2$&$845$ \\
\mmspr As                      & $ 8$&$27\times 10^{-6}$ & $ 4$&$50\times 10^{-7}$ & $  2$&$879$ \\
\mmspr Kr                      & $ 5$&$05\times 10^{-5}$ & $ 2$&$46\times 10^{-6}$ & $  2$&$666$ \\
\mmspr Mo                      & $ 1$&$88\times 10^{-4}$ & $ 8$&$00\times 10^{-6}$ & $  4$&$549$ \\
\mmspr Sn                      & $ 2$&$04\times 10^{-4}$ & $ 7$&$00\times 10^{-6}$ & $  4$&$351$ \\
\mmspr Xe                      & $ 6$&$30\times 10^{-5}$ & $ 1$&$96\times 10^{-6}$ & $  3$&$577$ \\
\mmspr Ba                      & $ 3$&$58\times 10^{-4}$ & $ 1$&$06\times 10^{-5}$ & $  4$&$301$ \\
\hline
\end{tabular}
\end{table}

\begin{figure*}
   \resizebox{\hsize}{!}{\includegraphics{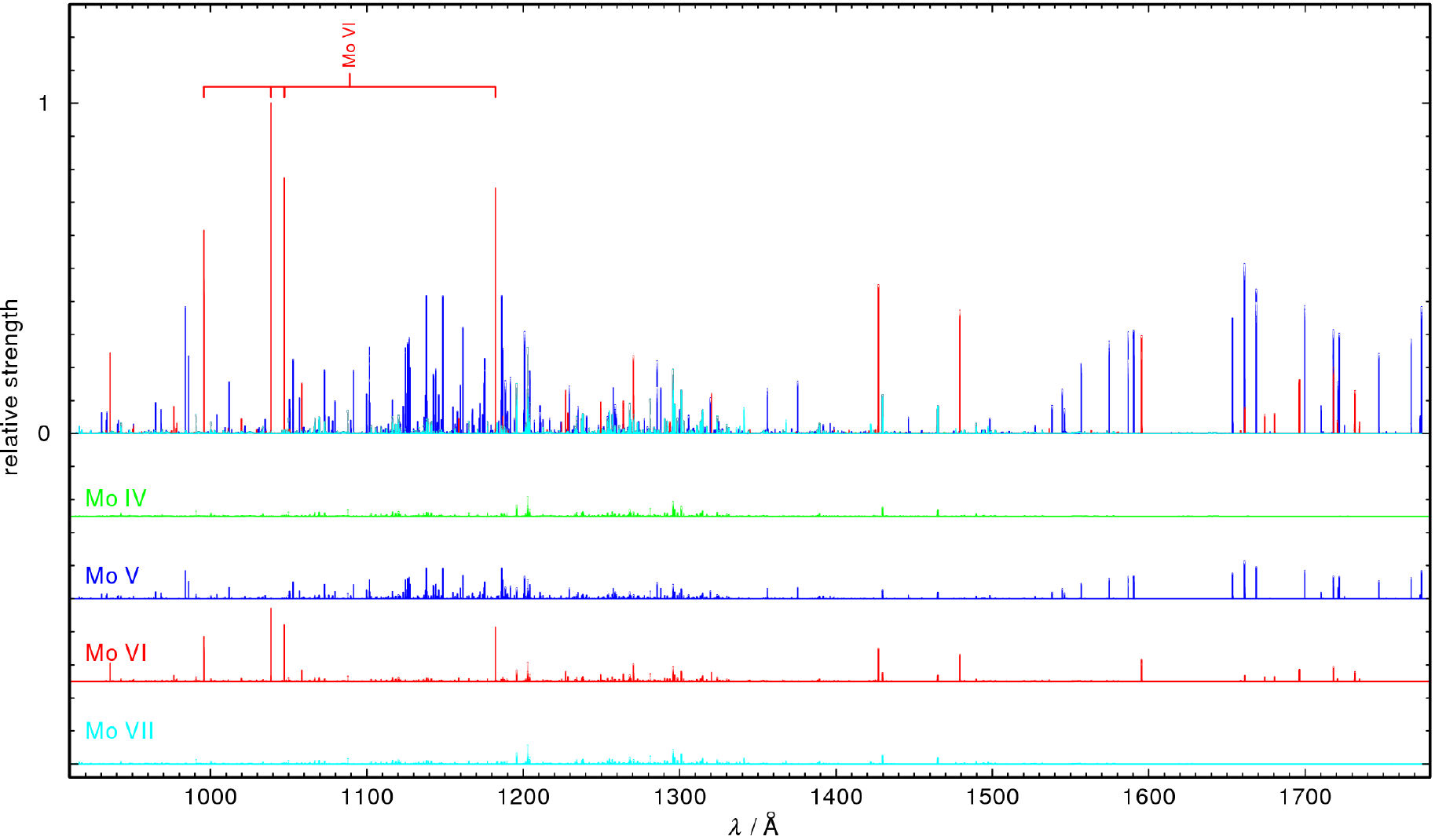}}
    \caption{Relative strengths of Mo lines calculated from our stellar-atmosphere model of \re. 
             Top panel: \ion{Mo}{iv-vii} lines, the four most prominent are \ion{Mo}{vi} lines that were identified by \citet{werneretal2012}
             are marked. Graphs 2 to 5 (from top to bottom): lines of individual \ion{Mo}{iv-vii} ions (intensities
             reduced by a factor of 0.22 compared to the top panel), respectively.
            }
   \label{fig:mospec}
\end{figure*}

Figure\,\ref{fig:mospec} shows the relative line strengths (normalized to that of \ionw{Mo}{vi}{1038.642}) of the
Mo lines in the synthetic spectrum of \re. We note that these strengths reduce if the spectrum is convolved with a Gaussian
to match the instruments' resolutions (Sect.\,\ref{sect:observation}).

\begin{figure*}
   \resizebox{\hsize}{!}{\includegraphics{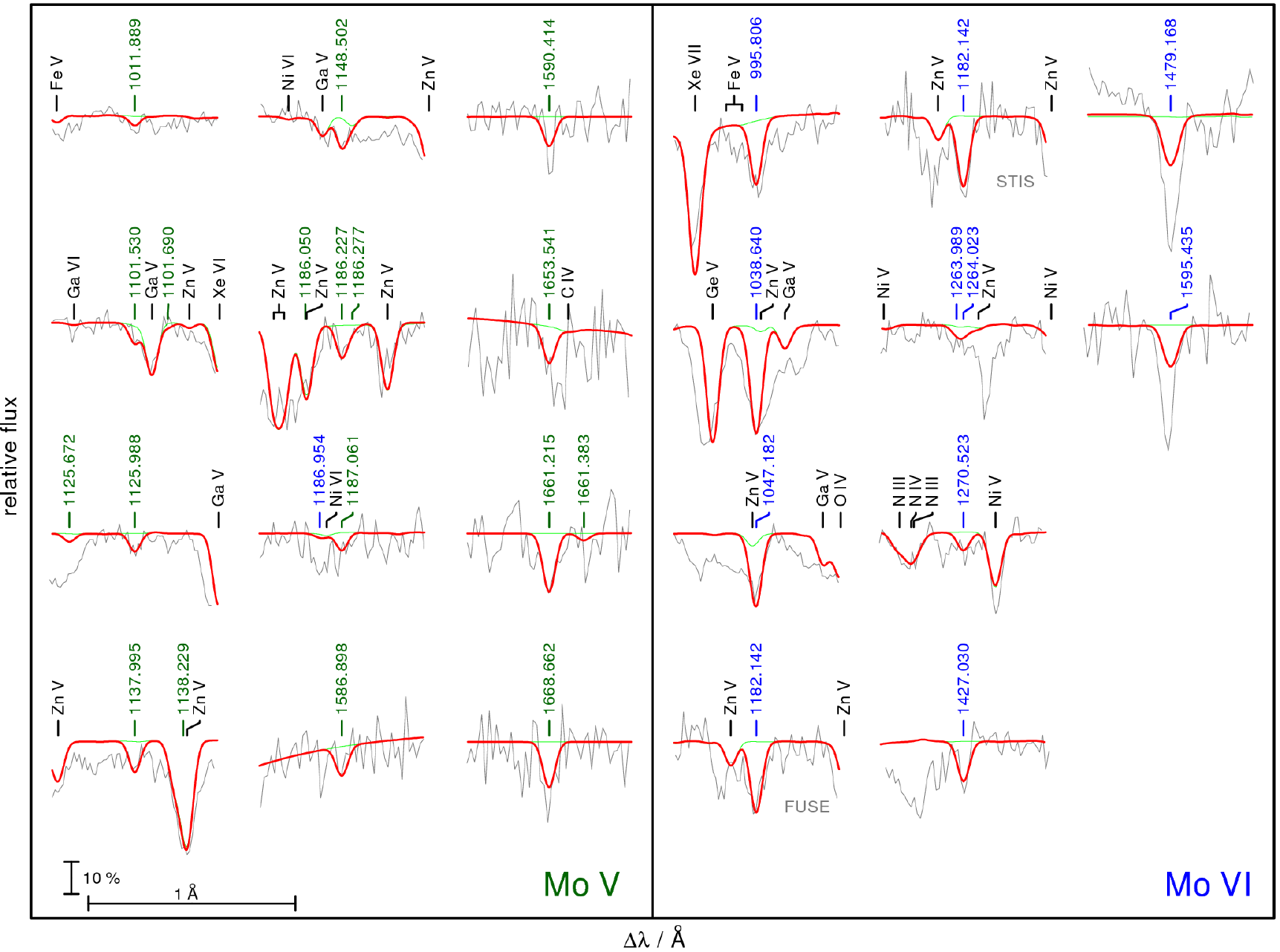}}
    \caption{\ion{Mo}{v}  lines (left  panel, marked with their wavelengths from Table~12 in \AA, green in the online version) and 
             \ion{Mo}{vi} lines (right panel, marked blue, wavelengths from Table~13)
             in the 
             FUSE (for lines at $\lambda < \mathrm{1188\,\AA}$) and 
             HST/STIS ($\lambda > \mathrm{1188\,\AA}$) observations of \re.
             The synthetic spectra are convolved with a Gaussian (full width at half maximum = FWHM = 0.06\,\AA)
             to simulate the instruments' resolutions.
             Other identified photospheric lines are marked in black.
             The thick red and thin green lines show a comparison with
             theoretical spectra of two models with and without Mo, respectively.  
             The vertical bar indicates 10\,\% of the continuum flux.
            }
   \label{fig:molines}
\end{figure*}

\begin{figure}
   \resizebox{\hsize}{!}{\includegraphics{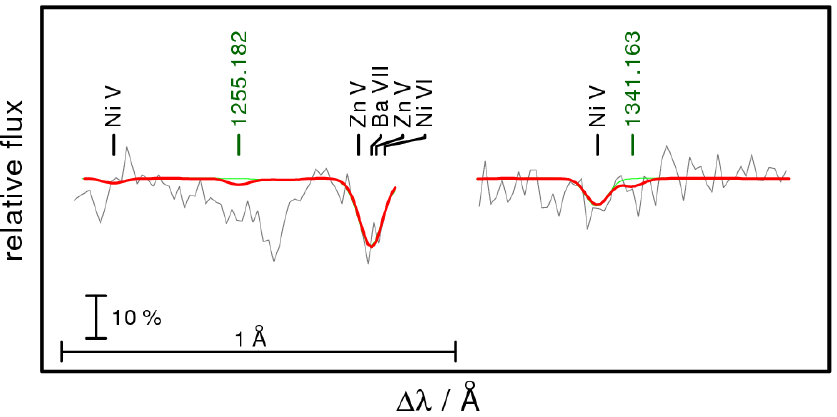}}
    \caption{Same as Fig.\,\ref{fig:molines}, for two \ion{Mo}{vii} lines.
            }
   \label{fig:mo_vii}
\end{figure}

In the FUSE and HST/STIS observations, we identified 12 \ion{Mo}{v} and nine \ion{Mo}{vi} lines
(Fig.\,\ref{fig:molines}). Their strengths were well reproduced at a Mo abundance of $1.88\times 10^{-4}$
(mass fraction), which is 35\,400 times the solar value \citep{grevesseetal2015}. Many more weak \ion{Mo}{v}
and \ion{Mo}{vi} lines are visible in the synthetic spectrum but they fade in the noise of the presently 
available observations.
The search for the strongest \ion{Mo}{iv} and \ion{Mo}{vii} lines (Fig.\,\ref{fig:mospec}) was not successful.
This was not unexpected, given the predicted weakness of the lines and the S/N of the data. Figure\,\ref{fig:mo_vii}
shows the two strongest \ion{Mo}{vii} lines of our model. We note that \ionw{Ba}{vii}{1255.520} dominates
the observed absorption around $\lambda  1255.5$\,\AA\ (Fig.\,\ref{fig:mo_vii}).

\begin{figure}
   \resizebox{\hsize}{!}{\includegraphics{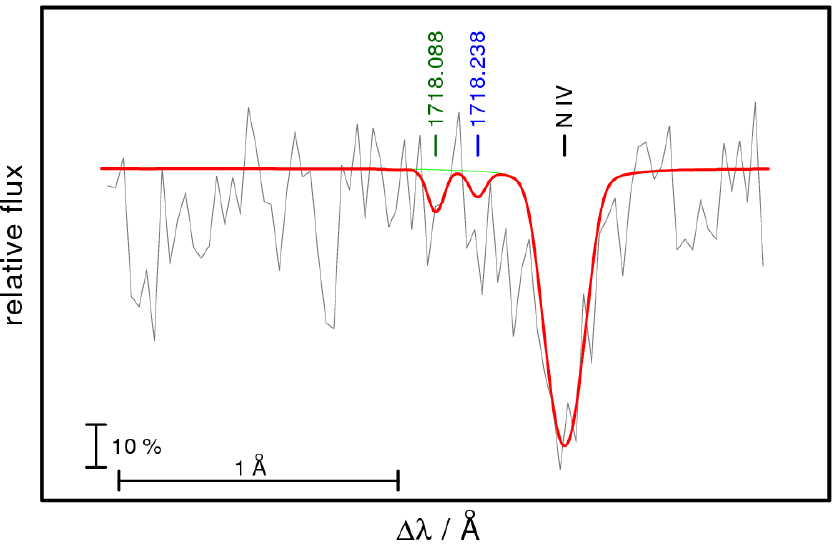}}
    \caption{Section of our HST/STIS observations of \re around \ionw{N}{iv}{1718.55}.
             The thick red and thin green lines show a comparison with
             theoretical spectra of two models with and without Mo, respectively.
             \ionw{Mo}{v}{1718.088} and \ionw{Mo}{vi}{1718.238} are marked.
            }
   \label{fig:molinesstis}
\end{figure}

The identification of Mo lines in the wavelength region $\lambda \ga 1700\,\mathrm{\AA}$ was strongly hampered by the lower
signal-to-noise (S/N) and resolution (only a fourth of the exposure time and 66\,\% of the resolving power
of the spectrum, compared to the region at $\lambda \la 1700\,\mathrm{\AA}$, see Sect.\,\ref{sect:observation}).
An example is shown in Fig.\,\ref{fig:molinesstis}. \ionw{Mo}{v}{1718.088} appears at comparable strengths of
other lines that were identified (Fig.\,\ref{fig:molines}). However, it is within the noise level of the
observation and has to be judged uncertain. \ionw{Mo}{vi}{1718.238} is weaker but a better observation would
allow an identification.

\subsection{\gb: Molybdenum}
\label{sect:mogb}

\begin{figure}
   \resizebox{\hsize}{!}{\includegraphics{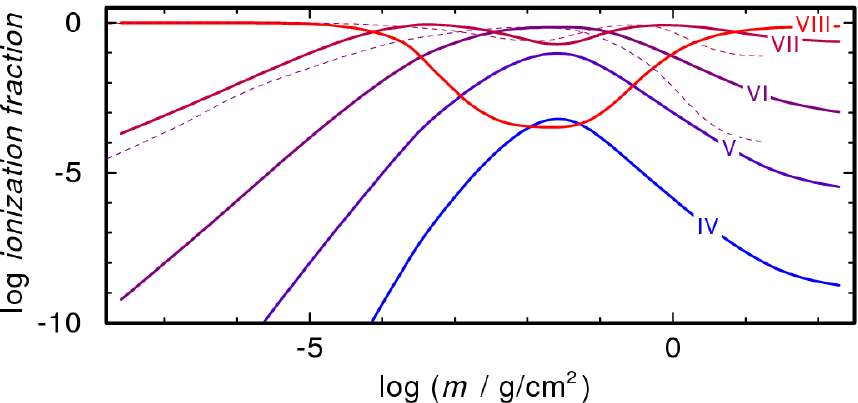}}
    \caption{Same as Fig.\,\ref{fig:ion_re} for \gb.
             For comparison, the dashed lines show the \ion{Mo}{vi+vii} ionization fractions in the \re model.
            }
   \label{fig:ion_gb}
\end{figure}

We added Mo into our atmosphere model (\Teffw{60\,000}, \loggw{7.6}) for \gb, which considers
H, He, C, N, O, Al, Si, P, S, Ca, Sc, Ti, V, Cr, Mn, Fe, Co, Ni, Zn, Ga, Ge, As, Mo, Sn, and Ba.
The abundances are given in Table\,\ref{tab:abgb}. \ion{Mo}{vi+vii} are the dominating ionization 
fractions in the line-forming region (Fig.\,\ref{fig:ion_gb}).

\begin{figure*}
   \resizebox{\hsize}{!}{\includegraphics{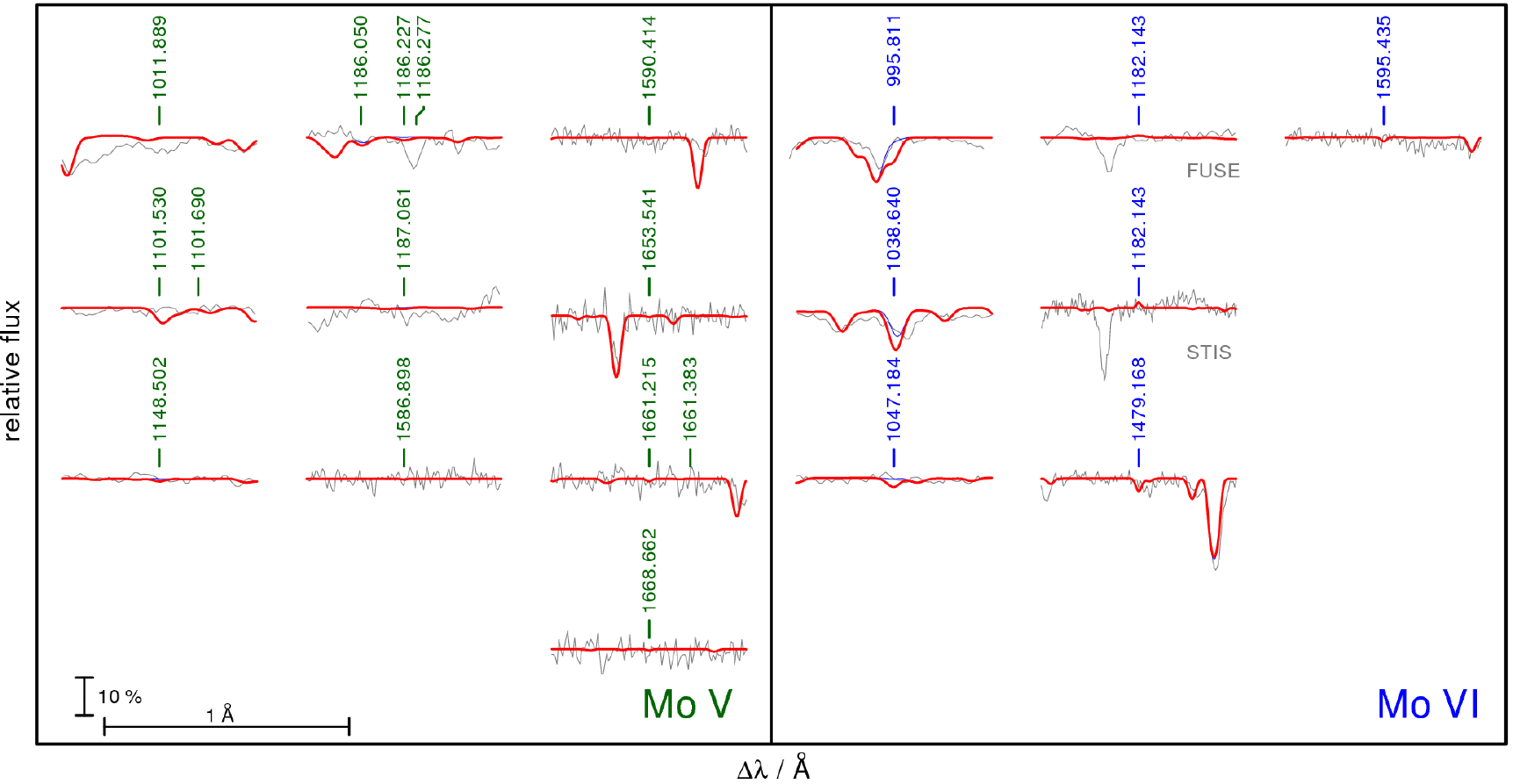}}
    \caption{Like Fig.\,\ref{fig:molines} for \gb. The thick red and thin blue spectra are calculated from
             models with photospheric Mo abundances of $5.3\times 10^{-6}$ and $5.3\times 10^{-7}$ (mass fractions, 
             about 1000 and 100 times the solar value), respectively.
            }
   \label{fig:molinesgb}
\end{figure*}

We performed a search for Mo lines in the observed spectra of \gb, analogous to that in Sect.\,\ref{sect:more}.
However, we did not identify any. Figure\,\ref{fig:molinesgb} shows a comparison of our synthetic
spectra to the observations. Since the HST/STIS spectrum of \gb (Sect.\,\ref{sect:observation}) is of excellent quality,
\ionw{Mo}{vi}{1469.168} gives a stringent upper abundance limit of $5.3\times 10^{-7}$ by mass \citep[about 100 times
solar,][]{grevesseetal2015}.

\begin{table}\centering 
  \caption{Same as Table\,\ref{tab:abre}, for \gb.}
\label{tab:abgb}
\setlength{\tabcolsep}{.4em}
\begin{tabular}{lr@{.}lr@{.}lr@{.}l}
\hline
\hline
\noalign{\smallskip}                                                                                          
                         & \multicolumn{2}{c}{Mass}   & \multicolumn{2}{c}{Number}  & \multicolumn{2}{c}{}                      \\
\cline{2-5}                     
\multicolumn{7}{c}{}                                                                                               \vspace{-5mm}\\
Element                  & \multicolumn{2}{c}{}       & \multicolumn{2}{c}{}        & \multicolumn{2}{c}{~~~~~[X]} \vspace{-2mm}\\
                         & \multicolumn{4}{c}{Fraction}                             & \multicolumn{2}{c}{}                      \\
\cline{1-7}                     
\noalign{\smallskip}                                                                                   
\mmspr H                       & $ 9$&$99\times 10^{-1}$ & $ 9$&$99\times 10^{-1}$ & $  0$&$132$ \\
\mmspr He                      & $<1$&$98\times 10^{-5}$ & $<5$&$00\times 10^{-6}$ & $<-4$&$099$ \\
\mmspr C                       & $ 6$&$31\times 10^{-6}$ & $ 5$&$30\times 10^{-7}$ & $ -2$&$574$ \\
\mmspr N                       & $ 2$&$08\times 10^{-6}$ & $ 1$&$50\times 10^{-7}$ & $ -2$&$522$ \\
\mmspr O                       & $ 1$&$90\times 10^{-5}$ & $ 1$&$20\times 10^{-6}$ & $ -2$&$479$ \\
\mmspr Al                      & $ 1$&$12\times 10^{-5}$ & $ 4$&$20\times 10^{-7}$ & $ -0$&$675$ \\
\mmspr Si                      & $ 5$&$29\times 10^{-5}$ & $ 1$&$90\times 10^{-6}$ & $ -1$&$099$ \\
\mmspr P                       & $ 1$&$54\times 10^{-6}$ & $ 5$&$00\times 10^{-8}$ & $ -0$&$579$ \\
\mmspr S                       & $ 5$&$72\times 10^{-6}$ & $ 1$&$80\times 10^{-7}$ & $ -1$&$733$ \\
\mmspr IG                      & $ 1$&$78\times 10^{-6}$ & $ 4$&$00\times 10^{-8}$ & $ -1$&$538$ \\
\mmspr Fe                      & $ 6$&$50\times 10^{-4}$ & $ 1$&$17\times 10^{-5}$ & $ -0$&$269$ \\
\mmspr Ni                      & $ 3$&$84\times 10^{-5}$ & $ 6$&$60\times 10^{-7}$ & $ -0$&$249$ \\
\mmspr Zn                      & $ 3$&$50\times 10^{-6}$ & $ 5$&$40\times 10^{-8}$ & $  0$&$304$ \\
\mmspr Ga                      & $ 2$&$56\times 10^{-6}$ & $ 3$&$70\times 10^{-8}$ & $  1$&$680$ \\
\mmspr Ge                      & $ 3$&$24\times 10^{-6}$ & $ 4$&$50\times 10^{-8}$ & $  1$&$155$ \\
\mmspr As                      & $ 3$&$71\times 10^{-7}$ & $ 5$&$00\times 10^{-9}$ & $  1$&$531$ \\
\mmspr Kr                      & $<1$&$09\times 10^{-6}$ & $<1$&$31\times 10^{-0}$ & $ <1$&$000$ \\
\mmspr Mo                      & $<5$&$33\times 10^{-7}$ & $<5$&$60\times 10^{-9}$ & $ <2$&$000$ \\
\mmspr Sn                      & $ 3$&$53\times 10^{-7}$ & $ 3$&$00\times 10^{-9}$ & $  1$&$589$ \\
\mmspr Xe                      & $<1$&$67\times 10^{-7}$ & $<1$&$28\times 10^{-9}$ & $ <1$&$000$ \\
\mmspr Ba                      & $ 4$&$00\times 10^{-6}$ & $ 2$&$94\times 10^{-8}$ & $  2$&$350$ \\
\hline
\end{tabular}
\end{table}

\subsection{\gb: Krypton  ($Z = 36$) and Xenon  ($Z = 54$)}
\label{sect:krxegb}

\begin{figure*}
   \resizebox{\hsize}{!}{\includegraphics{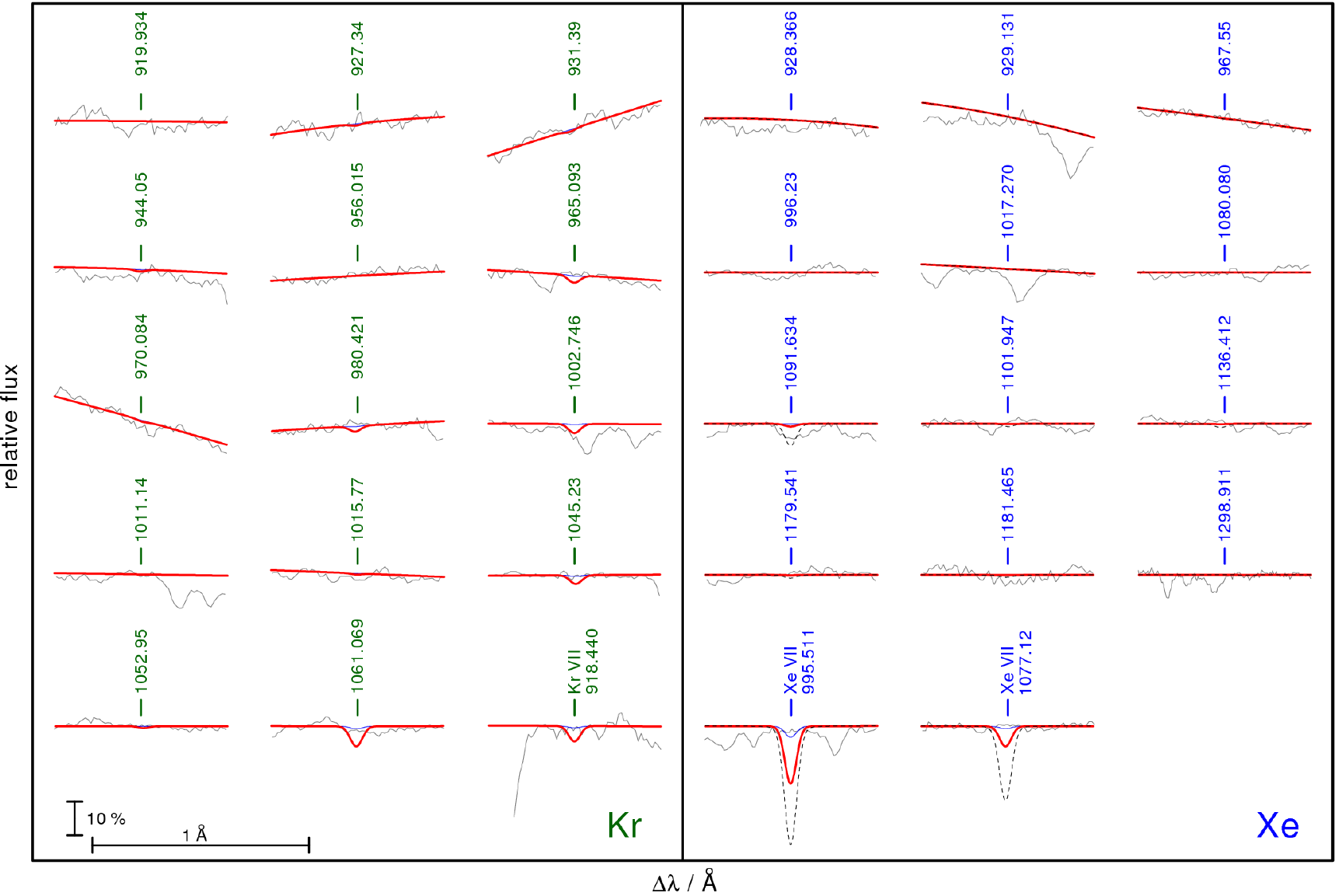}}
    \caption{Theoretical line profiles of Kr (left panel,  14 \ion{Kr}{vi} lines and 1 \ion{Kr}{vii} line) and 
                                           Xe (right panel, 12 \ion{Xe}{vi} and 2 \ion{Xe}{vii} lines) 
             compared to the observations of \gb.
             The spectra are calculated from models with photospheric abundances (mass fractions) of
             Kr: $1.1\times 10^{-5}$ (thick red,    100 times solar) and 
                 $1.1\times 10^{-6}$ (thin blue,     10 times solar) and of
             Xe: $1.7\times 10^{-6}$ (dashed black, 100 times solar),
                 $1.7\times 10^{-7}$ (thick red,     10 times solar), and 
                 $1.7\times 10^{-8}$ (thin blue,              solar).
           }
   \label{fig:krxe}
\end{figure*}

For \gb, we have determined a relatively high upper abundance limit (about 100 times solar) for Mo (Sect.\,\ref{sect:mogb}).
This is well in agreement with a factor of about $100 - 1000$ between the trans-iron element abundances in \re and \gb.
Like Mo, Kr, and Xe exhibit prominent lines in the UV spectra of \re, but not in those of \gb.
To investigate on their abundances, we individually included Kr and Xe in our \gb models and calculated
theoretical profiles for all Kr and Xe lines that were identified in \re \citep{werneretal2012,rauchetal2015xe}.
An upper Kr abundance limit of 10 times solar ($1.09 \times 10^{-6}$ by mass)
is determined from lines of \ion{Kr}{vi-vii} simultaneously (Fig.\,\ref{fig:krxe}).
In the case of Xe, the intersystem lines
\ionw{Xe}{vii}{995.51} (5s$^2$\,$^1$S -- 5s5p\,$^3$P$^\mathrm{o}$)
and
\ionw{Xe}{vii}{1077.12} (5s5p\,$^1$P$^\mathrm{o}$ -- 5p$^2$\,$^1$D)     
are very strong (Fig.\,\ref{fig:krxe}) and require an upper Xe abundance limit of solar to fade in the
noise of the observation. However, this may be strongly underestimated because of the rudimentary \ion{Xe}{vii} model atom 
presently provided by TMAD.
In that, only two \ion{Xe}{vii} lines with reliable oscillator strengths are known, namely,
0.245 for \ionw{Xe}{vii}{995.51} \citep{kernahanetal1980} and
0.810 for \ionw{Xe}{vii}{1077.12} \citep{biemontetal2007}.
Since, for the calculation of accurate NLTE occupation numbers of the atomic levels of an specific model ion,
reliable transition probabilities are mandatory for the complete ion,
the \ion{Xe}{vii} upper limit is regarded as uncertain.
This issue is out of the scope of this paper but will be investigated in detail immediately after new 
\ion{Xe}{iv-vii} transition probabilities become available. 
However, from the \ion{Xe}{vi} lines alone,
we achieve an upper limit of $1.7\times 10^{-7}$ (10 times the solar value, Fig.\,\ref{fig:krxe}).

\section{Impact of diffusion on trans-iron elements}
\label{sect:diffusion}

\begin{figure}
   \resizebox{\hsize}{!}{\includegraphics{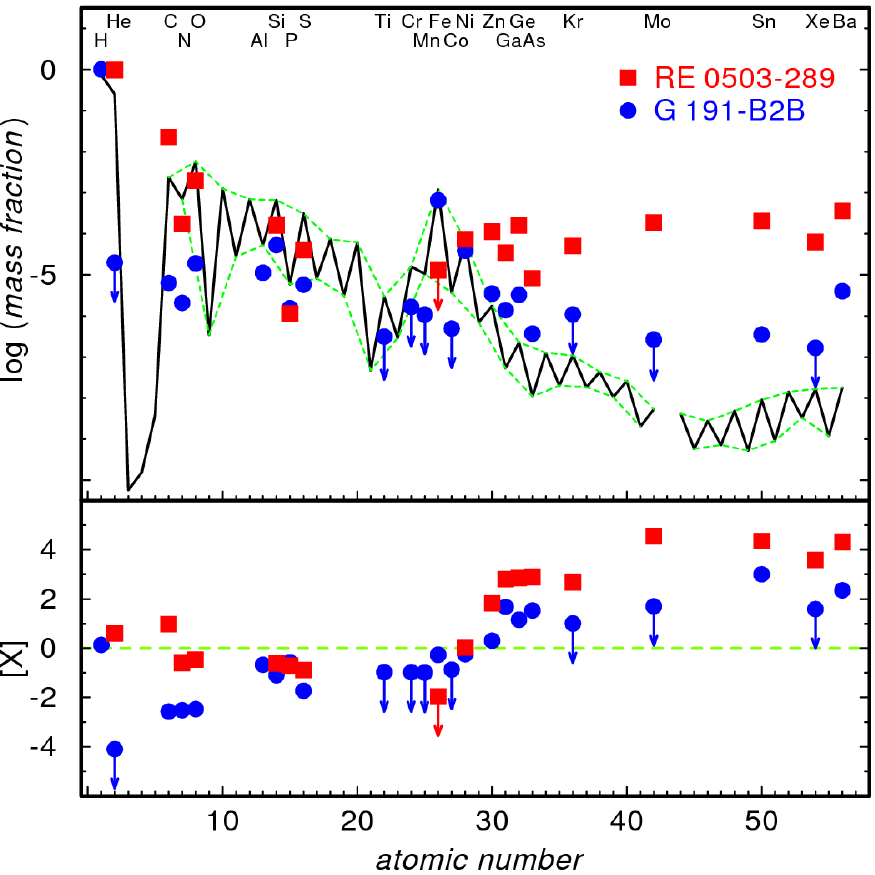}}
    \caption{Solar abundances \citep[thick line; the dashed green lines
             connect the elements with even and with odd atomic number]{asplundetal2009,scottetal2015a,scottetal2015b,grevesseetal2015}
             compared with the determined photospheric abundances of 
             \gb \citep[blue circles,][]{rauchetal2013} and 
             \re \citep[red squares,][and this work]{dreizlerwerner1996,werneretal2012,rauchetal2013,rauchetal2014zn,rauchetal2014ba,rauchetal2015xe,rauchetal2015ga}.
             Top panel: Abundances given as logarithmic mass fractions.
                        Arrows indicate upper limits.
             Bottom panel: Abundance ratios to respective solar values, 
                           [X] denotes log (fraction\,/\,solar fraction) of species X.
                           The dashed green line indicates solar abundances.
            }
   \label{fig:X}
\end{figure}

At almost the same \logg, \re has a significantly higher \Teff compared to that of \gb
(\Teffw{70\,000}, \loggw{7.5} vs\@. \Teffw{60\,000}, \loggw{7.6}, respectively). Thus, the much stronger
enrichment of the trans-iron elements in \re (Fig.\,\ref{fig:X}) may be the result of a more efficient 
radiative levitation. Therefore, we used the NGRT\footnote{New Generation Radiative Transport} code \citep{dreizlerwolff1999, schuhetal2002} 
to calculate diffusion models for both stars, 
using exactly the same model atoms for H, He, C, N, O, Ca, Sc, Ti, V, Cr, Mn, Fe, Co, Ni, Zn, Ga, Ge, As, Kr, Mo, Sn, Xe, and Ba,
which were used for our chemically homogeneous TMAP models.
For \re, H was formally included in the calculation, but its
abundance was fixed to $1.0\times 10^{-20}$. Therefore, its contribution to the background opacity is negligible.
Disregarding the fixed H abundance in \re, the diffusion models differ only in \Teff and \logg.

The TMAD model atoms for As and Sn are presently rather rudimentary, especially as only a very few oscillator strengths are
known. Restricting the radiative levitation calculation to include transitions with known oscillator strengths thus leads to an 
unrealistically small effect. We follow \citet[][]{rauchetal2013} and add all allowed line transitions
in our As and Sn model atoms, using default f-values of 1. Therefore, the results of our diffusion models for these elements should be 
regarded as preliminary.

\begin{figure}
  \resizebox{\hsize}{!}{\includegraphics{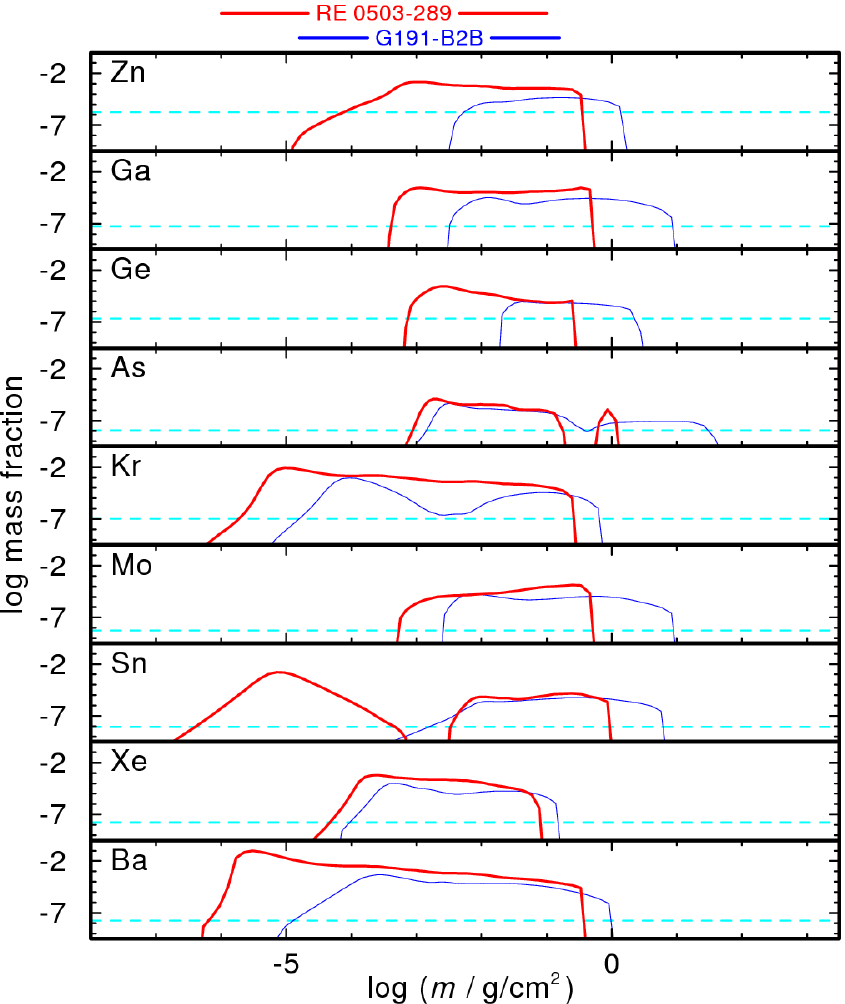}}
  \caption{Abundance profiles in our diffusion models for \gb (thin blue) and \re (thick red).
           The dashed, horizontal lines indicate solar abundance values.
           The formation regions of UV lines in both models are indicated at the top.
          } 
  \label{fig:depabund}
\end{figure}

Figure\,\ref{fig:depabund} shows the calculated depth-dependent abundance profiles for Zn, Ga, Ge, As, Kr, Mo, Sn, Xe, and Ba. 
All these are strongly overabundant in the line-forming regions (Fig.\,\ref{fig:depabund}).
The predicted abundance profiles suggest abundance enhancements in \re relative to \gb.
This is qualitatively in agreement with the abundance patterns in Fig.\,\ref{fig:X}, which were determined from our static TMAP models.
However, whether it is possible to reach 2-3\,dex should be demonstrated with advanced line-profile 
calculations in diffusive equilibrium. These would provide stringent constraints for suggested weak stellar winds \citep{chayeretal2005} 
and thin convective zones at the stellar surface \citep{chayeretal2015}, which might impact the interplay of radiative levitation and
gravitational settling.

\section{Results and conclusions}
\label{sect:results}

With our NLTE model-atmosphere package TMAP, 
we calculated new models for the DO-type white dwarf \re with molybdenum in addition.
The Mo model atoms were constructed with newly calculated \ion{Mo}{iv-vii} oscillator strengths.
We have unambiguously identified 12 \ion{Mo}{v} and nine \ion{Mo}{vi} lines in the observed high-resolution UV spectra of \re. 
The \ion{Mo}{v}\,/\,\ion{Mo}{vi} ionization equilibrium is well reproduced (Fig.\,\ref{fig:molines}).
We determined a photospheric abundance of 
$\log\,\mathrm{Mo} = -3.73 \pm 0.2$ (mass fraction $1.2 - 3.0\,\times\,10^{-4}$, 22\,500\,$-$\,56\,400 times the solar abundance).
In addition, we determined the arsenic and tin abundances and derived
$\log\,\mathrm{As} = -5.08 \pm 0.2$ ($0.5 - 1.3\,\times\,10^{-5}$, about 300\,$-$\,1200 times solar) and
$\log\,\mathrm{Sn} = -3.69 \pm 0.2$ ($1.3 - 3.2\,\times\,10^{-4}$, about 14\,300\,$-$\,35\,200 times solar).
These highly supersolar As, Mo, and Sn abundances agree well with the high abundances of other trans-iron elements in \re 
(Fig.\,\ref{fig:X}).

\gb does not exhibit Kr, Mo, and Xe lines in its UV spectrum. We investigated the strongest lines in the model
and found upper limits for the abundances of
Kr ($1.1\times 10^{-6}$, 10 times solar),
Mo ($5.3\times 10^{-7}$, 100 times solar), and
Xe ($1.7\times 10^{-7}$, 10 times solar).
Whether radiative levitation yields abundances of these elements that are consistent with observations should be demonstrated 
with advanced line-profile calculations in diffusive equilibrium, as depicted in Fig.\,\ref{fig:depabund}.
In addition, we determined the arsenic abundance and derived
$\log\,\mathrm{As} = -6.43 \pm 0.2$ ($2.3 - 5.9\,\times\,10^{-7}$, about 21\,$-$\,53 times solar).

The computation of reliable transition probabilities for \ion{Mo}{iv-vii} was a prerequisite
for the identification of Mo lines and the subsequent abundance determination. 
The hitherto known abundance pattern of \re (Fig.\,\ref{fig:X}) indicates that other yet
unidentified species should be detectable. Therefore, the
precise evaluation of their laboratory spectra, i.e., the measurement of
line wavelengths and strengths, as well as the determination of level energies and the 
subsequent calculation of transition probabilities are absolutely essential. 

The example of the arsenic abundance determination (Sect.\,\ref{sect:as}) had demonstrated that
state-of-the-art NLTE stellar-atmosphere models are mandatory for the precise spectral analysis 
of hot stars.

\begin{acknowledgements}
TR and DH are supported by the German Aerospace Center (DLR, grants 05\,OR\,1402 and 50\,OR\,1501, respectively).
The GAVO project had been supported by the Federal Ministry of Education and
Research (BMBF) 
at T\"ubingen (05\,AC\,6\,VTB, 05\,AC\,11\,VTB) and is funded
at Heidelberg (05\,AC\,11\,VH3).
Financial support from the Belgian FRS-FNRS is also acknowledged. 
PQ is research director of this organization.
We thank our referee, St\'ephane Vennes, for constructive criticism.
Some of the data presented in this paper were obtained from the
Mikulski Archive for Space Telescopes (MAST). STScI is operated by the
Association of Universities for Research in Astronomy, Inc., under NASA
contract NAS5-26555. Support for MAST for non-HST data is provided by
the NASA Office of Space Science via grant NNX09AF08G and by other
grants and contracts. 
This research has made use of 
NASA's Astrophysics Data System and
the SIMBAD database, operated at CDS, Strasbourg, France.
The TOSS service (\url{http://dc.g-vo.org/TOSS}) 
that provides weighted oscillator strengths and transition probabilities
was constructed as part of the
activities of the German Astrophysical Virtual Observatory.
\end{acknowledgements}

\bibliographystyle{aa}
\bibliography{27324}

\end{document}